%% file: Main.tex
\documentclass[%
 reprint,
superscriptaddress,
 amsmath,amssymb,
 aps,
]{revtex4-2}

\usepackage{verbatim}

\usepackage{graphicx}
\usepackage{dcolumn}
\usepackage{bm}
\usepackage[usenames,dvipsnames]{color}

\newcommand{\Ep}[1]{\left\langle#1\right\rangle_P}
\newcommand{\Eobs}[1]{\left\langle#1\right\rangle_{obs}}

\newcommand{\st}{\mathbf{s}_t}
\newcommand{\rt}{\mathbf{r}_t}
\newcommand{\vt}{\mathbf{v}_t}
\newcommand{\stau}{\mathbf{s}_{t-\tau}}
\newcommand{\Ft}{\mathbf{F}_{t-\tau}}
\newcommand{\Fti}{\bm{F}^{t-\tau}_i}

\newcommand{\Ftt}{\mathbf{F}_{t}}

\newcommand{\At}{\mathbf{A}_{t}}

\newcommand{\vtt}[1]{\mathbf{v}_{t_{#1}}}
\newcommand{\stt}[1]{\mathbf{s}_{t_{#1}}}

\newcommand{\rti}[1]{\bm{r}^t_{#1}}
\newcommand{\rtaui}[1]{\bm{r}^{t-\tau}_{#1}}
\newcommand{\vti}[1]{\bm{v}^t_{#1}}
\newcommand{\vtaui}[1]{\bm{v}^{t-\tau}_{#1}}
\newcommand{\staui}[1]{\bm{s}^{t-\tau}_{#1}}
\newcommand{\sti}[1]{\bm{s}^t_{#1}}

\newcommand{\rtauix}[1]{r^{t-\tau}_{#1,x}}
\newcommand{\rtauiy}[1]{r^{t-\tau}_{#1,y}}
\newcommand{\rtauiz}[1]{r^{t-\tau}_{#1,z}}

\newcommand{\Ht}{H[P|\stau]}

\newcommand{\Pt}{P(\mathbf{v}_t|\mathbf{s}_{t-\tau})}
\newcommand{\Pti}[1]{P(\mathbf{v}_{t_{#1+1}}|\mathbf{s}_{t_{#1}})}
\newcommand{\ft}[1]{f_{#1}(\mathbf{v}_t|\mathbf{s}_{t-\tau})}

\newcommand{\fki}[1]{f_{#1}^{kin} (\mathbf{v}_t)}

\newcommand{\lak}{\lambda_{kin}^{t,\tau}}
\newcommand{\laks}{\lambda_{kin}^{t,\tau, *}}
\newcommand{\lakt}{\lambda_{kin}^{t}}

\newcommand{\fwi}[2]{f_{#1}^{#2} (\mathbf{v}_t|\mathbf{s}_{t-\tau})}
\newcommand{\fw}{f_{w}(\mathbf{v}_t|\mathbf{s}_{t-\tau})}
\newcommand{\fww}{f_{w}^{t,\tau}}
\newcommand{\Qwi}[2]{\mathbf{Q}_{#1,#2}^{t,\tau}}
\newcommand{\lw}[1]{\lambda_{#1}^{t,\tau}}
\newcommand{\fwk}{f_{kin}^{t,\tau}}
\newcommand{\fwu}[1]{f_{#1}^{t,\tau}}
\newcommand{\gwi}[3]{g_{#1,#2}^{#3} (\stau;\eta_{#2})}
\newcommand{\uwi}[3]{\bm{u}_{#1,#2}^{#3} (\stau)}

\newcommand{\Qwt}[2]{\mathbf{Q}_{#1,#2}^{t}}
\newcommand{\lwt}[1]{\lambda_{#1}^{t}}
\newcommand{\Fwti}{\bm{F}^{t}_i}

\newcommand{\falii}[1]{f_{#1}^{ali} (\mathbf{v}_t|\mathbf{s}_{t-\tau})}

\newcommand{\lali}{\lambda_{ali}^{t,\tau}}
\newcommand{\nalii}{n^{t-\tau}_{i,ali}}
\newcommand{\Nalii}{N^{t-\tau}_{i,ali}}

\newcommand{\lalit}{\lambda_{ali}^{t}}

\newcommand{\naliit}{n^{t}_{i,ali}}

\newcommand{\frepi}[1]{f_{#1}^{rep} (\mathbf{v}_t|\mathbf{s}_{t-\tau})}

\newcommand{\larep}{\lambda_{rep}^{t,\tau}}
\newcommand{\nrepi}{n^{t-\tau}_{i,rep}}
\newcommand{\urepi}{\bm{u}_{i,j,rep}^{t-\tau}}

\newcommand{\fattri}[1]{f_{#1}^{att} (\mathbf{v}_t|\mathbf{s}_{t-\tau})}

\newcommand{\laattr}{\lambda_{att}^{t,\tau}}
\newcommand{\nattri}{n^{t-\tau}_{i,att}}
\newcommand{\uattri}{\bm{u}_{i,j,att}^{t-\tau}}

\newcommand{\fboi}[1]{f_{#1}^{bou} (\mathbf{v}_t|\mathbf{s}_{t-\tau})}

\newcommand{\labo}{\lambda_{bou}^{t,\tau}}
\newcommand{\uboi}{\bm{u}_{i,bou}^{t-\tau}}
\newcommand{\rboi}{\bm{r}_{i,bou}^{t-\tau}}

\newcommand{\nboi}{n^{t-\tau}_{i,bou}}
\newcommand{\nboixy}{n^{t-\tau}_{i,bou,xy}}
\newcommand{\nboiz}{n^{t-\tau}_{i,bou,z}}
\newcommand{\uboix}{u^{t-\tau}_{i,bou,x}}
\newcommand{\uboiy}{u^{t-\tau}_{i,bou,y}}
\newcommand{\uboiz}{u^{t-\tau}_{i,bou,z}}

\newcommand{\fexi}[1]{f_{#1}^{des} (\mathbf{v}_t|\mathbf{s}_{t-\tau})}

\newcommand{\laex}{\lambda_{des}^{t,\tau}}
\newcommand{\uexti}{\bm{u}_{i,des}^{t-\tau}}

\newcommand{\nexi}{n^{t-\tau}_{i,des}}

\newcommand{\lakle}{\lambda_{0,kin}^{t,\tau}}

\newcommand{\uaexle}{\bm{u}_{0,des}^{t,\tau}}

\newcommand{\uaexlexs}{u_{0,des,x}^{t,\tau,*}}
\newcommand{\uaexleys}{u_{0,des,y}^{t,\tau,*}}
\newcommand{\uaexlex}{u_{0,des,x}^{t,\tau}}
\newcommand{\uaexley}{u_{0,des,y}^{t,\tau}}

\newcommand{\LAt}{\mathbf{\Lambda}_{t,\tau}}

\newcommand{\LAtss}{\mathbf{\Lambda}^*_{t,\tau}}

\newcommand{\Zt}{Z(\LAt)}

\newcommand{\LAtt}{\mathbf{\Lambda}_{t}}
\newcommand{\Ztt}{Z(\LAtt)}

\newcommand{\Ptt}{P(\mathbf{v}_t|\mathbf{r}_{t})}

\newcommand{\Zti}[1]{Z(\mathbf{\Lambda}_{t_{#1+1},\tau})}

\begin{document}

\hyphenpenalty=10000
\tolerance=10000

\preprint{APS/123-QED}

\title{A General Statistical Mechanics Framework for the Collective Motion of Animals}%

\author{Jiacheng Cai}
 \email{jxcai@salisbury.edu}
 \affiliation{Department of Mathematical Sciences, Salisbury University, Salisbury, MD 21801, USA}

\author{Jianlong Zhang}
 \affiliation{Guangdong Key Laboratory for Innovation Development and Utilization of Forest Plant
Germplasm, College of Forestry and Landscape Architecture, South China Agricultural
University, Guangzhou, Guangdong 510642, China}

\author{Xuan Chen}
 \affiliation{Department of Biology, Salisbury University, Salisbury, MD 21801, USA}

\author{ Cai Wang}
 \email{wangcai@scau.edu.cn}
 \affiliation{Guangdong Key Laboratory for Innovation Development and Utilization of Forest Plant
Germplasm, College of Forestry and Landscape Architecture, South China Agricultural University, Guangzhou, Guangdong 510642, China}


\begin{abstract}
We propose a general statistical mechanics framework for the collective motion of animals. The framework considers the principle of maximum entropy, the interaction, boundary, and desire effects, as well as the time-delay effect. These factors provide the ability to describe and solve dynamic and non-equilibrium problems under this framework. We show that the Vicsek model, the social force model, and some of their variants can be considered special cases of this framework. Furthermore, this framework can be extended to the maximum caliber setting. We demonstrate the potential of this framework for model comparisons and parameter estimations by applying the model to observed data from a field study of the emergent behavior of termites. Finally, we demonstrate the flexibility of the framework by simulating some collective moving phenomena for birds and ants.

\end{abstract}

\maketitle


The collective motion of animals, as a fascinating natural phenomenon, is a complex topic that involves biology, physics, mathematics, and related fields \cite{Sumpter2006Review,Vicsek2012Review,Ouellette2021Models}. This topic has raised scientific interests in the theoretical and empirical studies of various species, such as schools of fish \cite{Partridge1982Fish, Hemelrijk2008Fish, Lopez2012Fish, Ioannou2012Fish, Jhawar2020Fish}, swarms of insects \cite{Couzin2003AntMill, Shiwakoti2011ant, Feinerman2018ant, Chandrae2021AntRaid, Lutz2021ant}, flocks of birds \cite{Ballerini2008birds,Bialek2012StatMech,Bialek2014MaxEnt, Cavagna2014birds,Cavagna2015birds,Mora2016birds,Cavagna2018Flock, Cristiani2021Birds}, and crowds of pedestrians \cite{Helbing1995SocialForce, Helbing1997human, Helbing2000Human, Festa2018KineticDO, QIN2018pedes, Sticco2021pedes}. Many theoretical models have focused on the microscopic description of each individual in a group during collective motion, such as the social force model \cite{Helbing1995SocialForce, Helbing1997human, Helbing2000Human, Hemelrijk2011SocialForce}, the Vicsek model \cite{Vicsek1995, Chate2008Vicsek}, and the Cucker-Smale model \cite{Cucker2007CS,Cucker2007CS2}. One of the challenges in theoretical modeling is to examine the various assumptions and to compare the fitness of the models based on the observed data. Statistical mechanics based on the principle of maximum entropy \cite{Jaynes1957, Bialek2012StatMech, Bialek2014MaxEnt, Cavagna2014birds, Mora2016birds} provides a bridge over the gap between theoretical modeling and empirical data. However, some of the maximum entropy models \cite{Bialek2012StatMech, Bialek2014MaxEnt} are based on the static setting, i.e., using the data at only one given point of time. This setting fails to capture the dynamics of some collective motion phenomena \cite{Cavagna2018Flock}. One of the approaches for dynamical analysis is the principle of maximum caliber \cite{Jaynes1980MaxCal,Presse2013MaxCal, Cavagna2014birds}. Specifically, for the case of flocks of birds, reference \cite{Cavagna2014birds} proposed a dynamical maximum entropy approach for the analysis of collective motion. In this approach, pairwise correlation functions between the birds' orientations at the same time point and between two consecutive time points were analyzed and summed up over the entire trajectory. It is equivalent to analyzing the resultant interaction forces for each individual. It remains unclear how to extend the framework to analyze the component repulsion, attraction, boundary, and desire (the will of the animal) forces for a more general setting.

In this work, we present a general statistical mechanics framework for describing the collective motion of a group of animals. With consideration of the maximum entropy principle, the interaction, boundary, and desire effects, as well as the time-delay effect of information delivery among the group \cite{Erban2016Delay,Szwaykowska2016Delay,Chen2021Delay,Cristiani2021Birds}, this framework provides a ``bottom to top'' analytical bridge between empirical data and theoretical models, as well as a simulation tool.  We show that some common models, such as the Vicsek model, the social force model, and some of their variants, are special cases of this general framework. Furthermore, this framework can be extended to the maximum caliber setting.
With this general framework, we apply model analysis to the empirical data from a field study of emergent behavior of termites, \textit{Odontotermes formosanus} \cite{Xiong2018termites}. Demonstrations are performed for model comparisons and parameter estimations are made based on maximum likelihood. We find that the Vicsek model assumption with adjustments has a larger likelihood compared with another particular model. In addition, we select the best estimation of delay time from several candidates in this particular field study case based on likelihood values. For the general modeling framework, we also propose some particular models for simulating the collective behaviors of death mills \cite{Couzin2003AntMill} and mass raids of army ants \cite{Chandrae2021AntRaid}, as well as birds flocking and escaping from predators \cite{Hemelrijk2011SocialForce,Youtube2020}.

\section{\label{sec:result} Methods and Results}

\subsection{\label{subsec:model} The model}

The foundation of this modeling framework is the principle of maximum entropy \cite{Jaynes1957}, which is widely used in the fields of statistical physics and information theory. The theory states that, given the constraints from the observed data, the best choice of the probability distribution models is the choice that has the maximum entropy. Several  research studies have extended such ideas to collective behaviors, such as flocks of birds \cite{Bialek2012StatMech, Bialek2014MaxEnt} or networks of neurons \cite{Tkacik2014}. This principle selects the probability density function that has no assumptions except for the observational constraints. Therefore, the choice of observational constraints is the most critical part of the modeling.

Consider a system of $N$ animals whose motion state at time $t$ can be described by a set of vectors $\st=(\rt, \vt)$, where $\rt=(\rti{1},\rti{2},\cdots, \rti{N})$  and $\vt=(\vti{1},\vti{2},\cdots, \vti{N})$ are the positions and velocities of the animals at time $t$, respectively. $\bm{r}^t_i$ and $\bm{v}^t_i$ can be either two-dimensional ($d=2$) or three-dimensional ($d=3$) vectors, depending on the study interest. We also add the notation $\sti{i}=(\rti{i},\vti{i})$.
We define the ``dot product'' for this type of data structures: consider $\mathbf{a}=(\bm{a}_1,\dots,\bm{a}_N)$ and $\mathbf{b}=(\bm{b}_1,\dots,\bm{b}_N)$, where $\bm{a}_i=(a_{i1},\dots,a_{id}), \bm{b}_i=(b_{i1},\dots,b_{id})$ are d-dimensional vectors. The ``dot product'' of $\mathbf{a}$ and $\mathbf{b}$ is given by

\begin{equation*}
\mathbf{a}\cdot\mathbf{b}=\sum_{i=1}^N\sum_{j=1}^d a_{ij}b_{ij}.
\end{equation*}

We assume that a delayed reaction exists in the system, i.e., $\vti{i}$ reacts to the conditions of time $t-\tau$ for some $\tau>0$. If the state $\stau$ is known, we are interested in some measurements at time $t$ that can be expressed as $k$ functions: $\ft{1},\ft{2},\cdots, \ft{k}$. Our goal is to find a probability density function $\Pt$ such that:
\begin{itemize}
    \item[1.] The entropy
    \begin{equation}
    \Ht=-\int\Pt\ln\Pt d\vt \label{eqn:C1}
    \end{equation}
    is maximized.
    \item[2.] The expectation constraints
    \begin{equation}
    	\Ep{\ft{w}} = \Eobs{\ft{w}} \label{eqn:C2}
    \end{equation}
    are satisfied for all $w$. $\Ep{\cdot}$ is the expectation under $P$ and  $\Eobs{\cdot}$ is the average based on the observed data.
\end{itemize}

Considering the validation and flexibility of the Vicsek model, the social force model, and their variants, we adopt the concept of the ``forces'', but we focus on their directions instead of their specific expressions. In addition, we express these directions as the functions of $\stau$. With the ``kinetic energy'', the effects of the alignment, repulsion, attraction, boundary, and ``desire'' (the will of each animal) factors, the corresponding maximum entropy distribution is

\begin{eqnarray}
&\Pt = \frac{1}{\Zt}\exp\left[-\frac{1}{2}\lak\vt\cdot\vt + \Ft\cdot\vt\right] \nonumber\\
&=\left(\frac{2\pi}{\lak}\right)^{-\frac{dN}{2}} \exp\left[-\frac{\lak}{2}\left(\vt-\frac{\Ft}{\lak}\right)\cdot\left(\vt-\frac{\Ft}{\lak}\right)\right],\nonumber\\ \label{eqn:P0}
\end{eqnarray}
where
\begin{eqnarray}
    \Zt &= &\int\exp\left[-\frac{1}{2}\lak\vt\cdot\vt + \Ft\cdot\vt\right]d\vt \nonumber\\
        &= &\sqrt{\left(\frac{2\pi}{\lak}\right)^{dN}}\exp\left[\frac{1}{2\lak}\Ft\cdot\Ft\right] \label{eqn:Zt2}
\end{eqnarray}
is the normalization factor.

$\lak$ can be considered the measurement of the ``average inertia'' of the system from time $t-\tau$ to $t$; $\Ft$ is an $N-$dimensional ``vector'' whose $i-$th component is a $d-$dimensional vector given by
\begin{eqnarray}
\Fti&=&\lali\sum_{j\in\nalii}\gwi{i,j}{ali}{t-\tau}\vtaui{j} \nonumber\\
&+&\larep\sum_{j\in\nrepi}\gwi{i,j}{rep}{t-\tau}\uwi{i,j}{rep}{t-\tau} \nonumber\\
&+&\laattr\sum_{j\in\nattri}\gwi{i,j}{att}{t-\tau}\uwi{i,j}{att}{t-\tau} \nonumber\\
&+&\labo\gwi{i}{bou}{t-\tau}\uwi{i}{bou}{t-\tau} \nonumber\\
&+&\laex\gwi{i}{des}{t-\tau}\uwi{i}{des}{t-\tau}, \label{eqn:Ft}
\end{eqnarray}
where $\urepi$ is the ``repulsion'' unit vector with the direction from $\rtaui{j}$ to $\rtaui{i}$ and $\uattri$ is the ``attraction'' unit vector with the direction from $\rtaui{i}$ to $\rtaui{j}$.  $\nalii, \nrepi, \nattri$ are the alignment, repulsion, and attraction neighbors, respectively, of the $i-$th animal at time $t-\tau$. If the $j-$th animal is within the corresponding neighbors at time $t-\tau$, then it has the corresponding effect to the $i-$th animal at time $t-\tau$, and this effect is reacted by the $i-$th animal at time $t$.
Similarly, we can define $\nboi$ and $\nexi$ as the index set of the individuals that are affected by the ``boundary'' and ``desire'' effect, respectively, at time $t-\tau$ and respond at time $t$.
$\uboi$ is the ``boundary'' unit vector with the direction from the closest boundary point to $\rtaui{i}$, and $\uexti$ is the ``desire'' unit vector representing the direction of the will of the $i-$th animal. $\uboi$ and/or $\uexti$ terms are set to zero vectors when $i$ is not within the corresponding $n$ terms.
It should be pointed out that the definition of neighbors can be either metric \cite{Vicsek1995} or topological \cite{Ballerini2008} in our framework.

Finally, we consider the strengths of the corresponding effects to the $i-$th animal from time $t-\tau$ to time $t$. We assume that the strength of each effect consists of two parts: first, the part that is a constant over the whole system at time $t-\tau$, described by the $\lambda$ terms; second, the part that is specified as a particular function of $\staui{i}$ (and $\staui{j}$ for interaction effects), described by the $g$ terms. The $\eta$ terms are parameters of the $g$ functions.
For a specified model, the formats of the $n$ terms, the formulas of the $g$ terms, and the value of the $\eta$ terms are given. We emphasize that we do not require the formats/formulas/values of the $n/g/\eta$ terms to remain unchanged over time, as long as the formats/formulas/values are known at time $t-\tau$. This provides the ability to describe and solve dynamic and non-equilibrium problems under this framework.

At this point, we have three sets of parameters in this framework: $\tau$, $\eta$, and $\lambda$. We can choose reasonable or empirical values as candidates for $\tau$ and $\eta$, and select the values with the best likelihood. For the $\lambda$ terms, we can use maximum likelihood estimations.

Take $\vt, \Ft$ as $dN-$dimensional vectors. The target distribution (\ref{eqn:P0}) is a multivariate normal distribution, the mean and covariance of which are given by
\begin{eqnarray}
\bm{\mu_{t,\tau}} &=& \frac{1}{\lak}\Ft, \label{mu}\\
\bm{\Sigma_{t,\tau}} &=& \frac{1}{\lak}\mathbf{I} \label{var},
\end{eqnarray}
where $\mathbf{I}$ is the $dN-$dimensional identity matrix. Given the $g$ and $\eta$ terms and the state information $\stau$, the distribution is uniquely determined by the $\lambda$ parameters $\LAt=\left(\lak,\lali,\larep,\laattr,\labo,\laex\right)$. It should be pointed out that $\LAt$ are the Lagrange multipliers for the following measurement functions stated in the condition (\ref{eqn:C2}), respectively:

\begin{eqnarray}
    \hspace{-3ex}\ft{kin} &=& \frac{1}{2N}\vt\cdot\vt; \\
    \hspace{-3ex}\ft{ali} &=& -\frac{1}{N}\sum_{i=1}^N\sum_{j\in\nalii}\gwi{i,j}{ali}{t-\tau}\vtaui{j}\cdot\vti{i}; \\
    \hspace{-3ex}\ft{rep} &=& -\frac{1}{N}\sum_{i=1}^N\sum_{j\in\nrepi}\gwi{i,j}{rep}{t-\tau}\uwi{i,j}{rep}{t-\tau}\cdot\vti{i}; \nonumber\\
    \hspace{-3ex}&&\\
    \hspace{-3ex}\ft{att} &=& -\frac{1}{N}\sum_{i=1}^N\sum_{j\in\nattri}\gwi{i,j}{att}{t-\tau}\uwi{i,j}{att}{t-\tau}\cdot\vti{i}; \nonumber\\
    \hspace{-3ex}&&\\
    \hspace{-3ex}\ft{bou} &=& -\frac{1}{N}\sum_{i=1}^N\gwi{i}{bou}{t-\tau}\uwi{i}{bou}{t-\tau}\cdot\vti{i}; \\
    \hspace{-3ex}\ft{des} &=& -\frac{1}{N}\sum_{i=1}^N\gwi{i}{des}{t-\tau}\uwi{i}{des}{t-\tau}\cdot\vti{i}.
\end{eqnarray}
They are the measurements of average reactions of $\vt$ at time $t$ to the corresponding effects from time $t-\tau$ over the whole system.

The derivation of the framework is inspired by the ideas of references \cite{Jaynes1957,Bialek2012StatMech} and is explained in details in the SI Appendix A.

\subsection{\label{subsec:other}Relationships with other models}

Since the general framework only specifies the directions of the effects from $\stau$ to $\vt$ and the detailed formats of the effects are not specified yet in the $g$ terms, this framework has a large flexibility to be adjusted. Many widely used models can be considered special cases of our framework (See SI Appendix B).

\begin{itemize}
\item When there is only one animal in the system and all other effects do not exist, we define $\nalii=\{i\}$ and set $\lali=\lak$. The normal distribution (\ref{eqn:P0}), along with (\ref{mu},\ref{var}), is equivalent to the following discrete dynamic model:

\begin{eqnarray*}
\rti{i}&=&\rtaui{i} + \vtaui{i}\tau,\\
\vti{i}&=&\vtaui{i}+\bm{\epsilon}_i^{t,\tau},
\end{eqnarray*}

where $\bm{\epsilon}_i^{t,\tau}$ is the white noise term that follows a d-dimensional normal distribution with the mean of $\bm{0}$ and the covariance of $1/\lak\mathbf{I}$. This is the law of inertia with the adjustment of randomness.

\item When we only consider the alignment effect and set $\lali=\lak$,  $\gwi{i,j}{ali}{t-\tau}=1/\Nalii$, where $\Nalii$ is the number of elements in $\nalii$, the normal distribution (\ref{eqn:P0}), along with (\ref{mu},\ref{var}), is equivalent to the following discrete dynamic model:

\begin{eqnarray*}
\rti{i}&=&\rtaui{i} + \vtaui{i}\tau,\\
\vti{i}&=&\sum_{j\in\nalii}\vtaui{j}/\Nalii+\bm{\epsilon}_i^{t,\tau}.
\end{eqnarray*}

This is equivalent to the Vicsek model with the relaxation of the fixed speed constraint.

\item For the general case, the model (\ref{eqn:P0}), along with (\ref{mu},\ref{var}), can be rewritten as
\begin{eqnarray}
\rti{i}-\rtaui{i}&=&  \vtaui{i}\tau, \label{eqn:r}\\
\vti{i}-\vtaui{i}&=& (\Fti/\lak-\vtaui{i})+\bm{\epsilon}_i^{t,\tau}. \label{eqn:v}
\end{eqnarray}

$(\Fti-\lak\vtaui{i})/\tau$ is equivalent to the discrete approximation of the interaction forces and the forces from the boundary and desire effects. We present the mapping between the social force model and our framework in the SI Appendix B.

\item Consider a sequence of times $t_0, t_1, \dots, t_M$ where $t_i=t_0+i\tau, i=1,\dots, M$. Given the initial distribution $P(\mathbf{v}_{t_0})$, the joint distribution of $(\vtt{0},\vtt{1},\dots,\vtt{M})$ can be derived as
\begin{eqnarray}
    &\qquad P(\vtt{0},\vtt{1},\dots,\vtt{M})=P(\mathbf{v}_{t_0})\prod_{i=0}^{M-1} \Pti{i}, \nonumber\\
    &\label{eqn:Pc}
\end{eqnarray}
where $\Pti{i}$ is derived from the maximum entropy principle (\ref{eqn:P0}). This distribution maximizes the caliber for the Markov setting with additional assumptions \cite{Ge2012MaxCalMC, Presse2013MaxCal, Dixit2015MaxCalMC}. See more details in the SI Appendix B.

\end{itemize}

\subsection{\label{subsec:termite} The emergent behavior of termites: an example}

Given the $g$ and $\eta$ terms, as well as the data of $\stau$ and $\st$, we can obtain the maximum likelihood estimators $\LAt{}^*=\left(\lak{}^*,\lali{}^*,\larep{}^*,\laattr{}^*,\labo{}^*,\laex{}^*\right)$ (See more details in SI Appendix C). In addition, we can find an estimation of the delay time $\tau^*$ if multiple candidates of $\tau$ and their corresponding $\stau$ are available. We select the $\tau$ and the corresponding $\stau$ that maximize the log likelihood of equation (\ref{eqn:P0}) among the candidates. This framework can be extended for comparisons of different models, as long as the differences of the models are in the selections of the neighbors and $g$ terms, which are just functions of $\stau$.

\begin{figure}
\centering
\begin{tabular}{cc}
\includegraphics[width=0.4\linewidth]{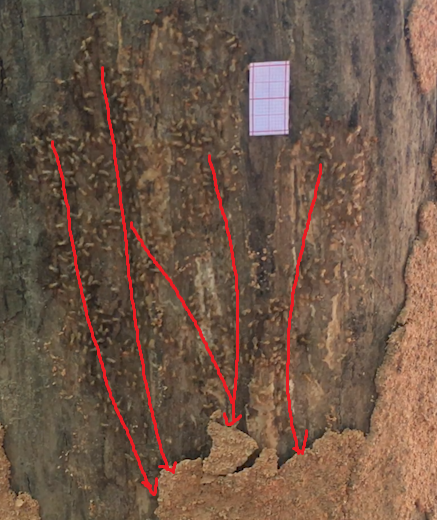}&
\includegraphics[width=0.4\linewidth]{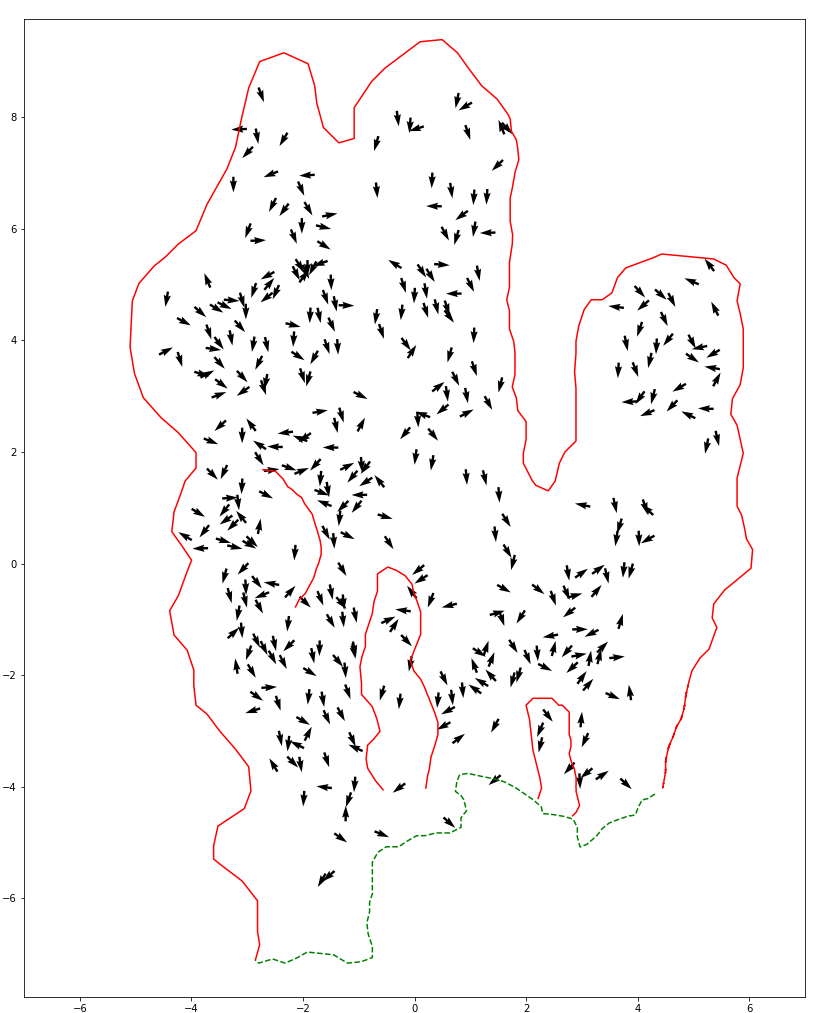}\\
A & B
\end{tabular}
\caption{(A) Snapshot of the termite activities (N=384) right after the sheeting is rapidly broken. The majority of the termites escape to the wide exit in the bottom within 10 seconds at the direction of the red arrows. (B) Instantaneous normalized vector velocities of all the agents in this snapshot.}
\label{fig:F1}
\end{figure}

As an example, we apply the above framework to the observed data from an emergent event of a colony of termites. This type of event has been reported in reference \cite{Xiong2018termites}. The data were collected in an arboretum in Guangzhou, China in 2019. The activities of termites are checked by observing the sheeting and mud tubes on tree trunks. After a camera is set to begin recording video, the sheeting is rapidly broken using a small rake and the termites escape to the exit below (SI Video 1, FIG. \ref{fig:F1}). Two-dimensional data for the positions and motions are tracked and collected by Tracker (version 5.1.3), a video analysis and modeling software program \cite{Tracker}. We choose $t=0.133s, 1.133s, 2.133s, \dots, 9.133s$, and $\tau=0.033s, 0.067s, 0.100s, 0.133s$. For each combination of $(t,\tau)$, we collect the data for $\st$ and $\stau$. The boundary data are also collected.

\begin{figure}
\centering
\includegraphics[width=\linewidth]{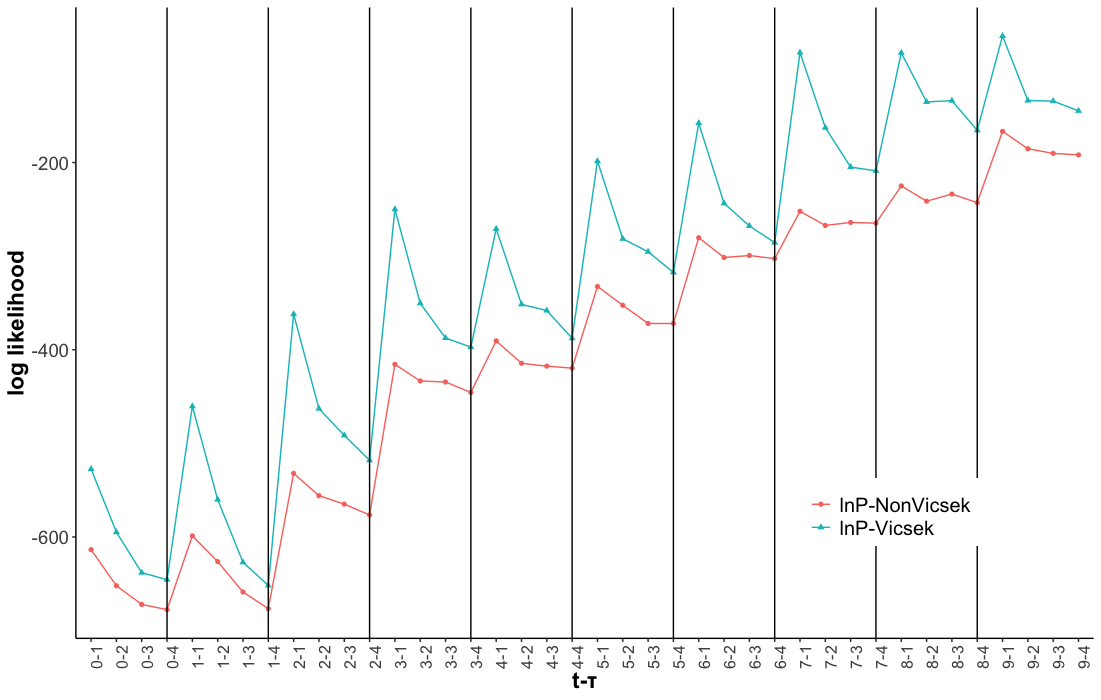}
\caption{Log likelihood of models with and without the Vicsek setting for all combinations of $(t,\tau)$. The x label of format $a-b$ indicates $a$ as the first digit of $t$ and $b=\tau/0.033$.}
\label{fig:F2}
\end{figure}

\begin{figure*}
\centering
\begin{tabular}{cc}
\includegraphics[width=0.4\linewidth]{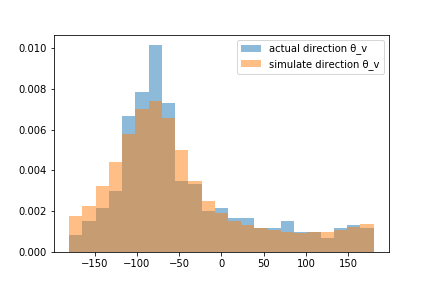}&
\includegraphics[width=0.4\linewidth]{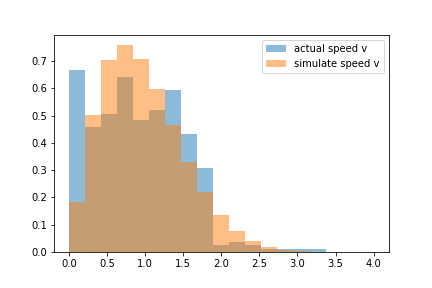}\\
A & B\\
\includegraphics[width=0.4\linewidth]{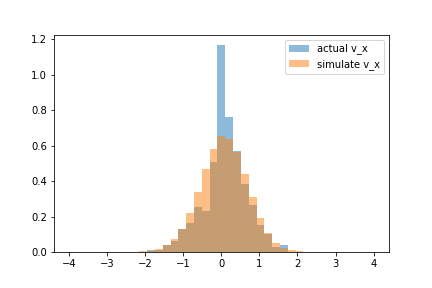}&
\includegraphics[width=0.4\linewidth]{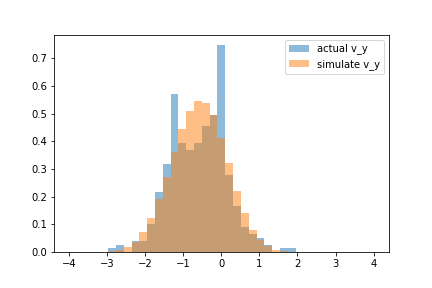}\\
C & D\\
\end{tabular}
\caption{
Comparisons of probability density histograms between the actual observed data and the 1000 simulations for all termites (at $t=0.133s$ with $\tau^*=0.033s$). Simulations are based on the model with the Vicsek setting. Figures show the comparisons for : (A) velocity angles in the unit of degree; (B) speed in the unit of cm/s; (C) projection of velocities on the $x$ direction in the unit of cm/s; (D) projection of velocities on the $y$ direction in the unit of cm/s.}
\label{fig:F3}
\end{figure*}

Worker termites have no visual sense \cite{Krishna2012termite}. In a short escape time period, we assume that the motion interactions are based on physical contacts, and we define $\nalii=\{j|\hspace{1ex}\lVert\rtaui{i}-\rtaui{j}\rVert\leq l_0\}$ and  $\nrepi=\{j|\hspace{1ex} 0<\lVert\rtaui{i}-\rtaui{j}\rVert\leq 0.5l_0\}$, where $l_0$ is the average body length of termites in this study. We assume that either there is no attraction effect, or the attraction effect is included in the desire effect of escaping with the direction $\uexti=(0,-1)$ for all $i,t,\tau$. We assume that $\gwi{i,j}{rep}{t-\tau}=\gwi{i}{bou}{t-\tau}=\gwi{i}{des}{t-\tau}=1$ for all $i,j,t,\tau$. For the alignment effect, we compare the log likelihood based on equation (\ref{eqn:P0}) between the model with $\gwi{i,j}{ali}{t-\tau}=1$ (non-Vicsek settings) and the model with $\gwi{i,j}{ali}{t-\tau}=1/\Nalii, \lali=\lak$ (Vicsek settings) at all combinations of $(t,\tau)$. The results in FIG. \ref{fig:F2} show that the models with $\tau^*=0.033s$ have the largest likelihood among the four candidates of $\tau$ for all different time $t$. The model with the Vicsek setting has larger likelihoods than the non-Vicsek setting for all combinations of $(t,\tau)$ in this experiment. Finally, we run 1000 simulations for each termite based on the model (\ref{eqn:P0}) with the Vicsek setting and the corresponding maximum likelihood estimators $\LAt^*$ for all $t$ with $\tau^*=0.033s$. We compare the probability density histograms of the actual observed data and the simulation results for the speeds $v$, the $x$ and $y$ projections of the velocities, and the velocity angles $\theta$. The results for $t=0.133s$ are presented in FIG. \ref{fig:F3}, which shows the validation of the theoretical model for the observed data. Other results and a detailed explanation of methods can be found in the SI Appendix C, D.

The purpose of this section is to present what can be computed from this framework. Further interpretations and inferences of the results are left for future studies.

\subsection{\label{subsec:simu} Simulations}

With the flexibilities of $g$ terms, the formats of the ``effects'', and $n$ terms, the effect neighbors, this general modeling framework has a strong ability to simulate several animal collective behaviors. In this section, we demonstrate the simulations of the death mill \cite{Couzin2003AntMill} and mass raids \cite{Chandrae2021AntRaid} of army ants, as well as birds flocking and escaping from predators \cite{Hemelrijk2011SocialForce,Youtube2020}. Detailed settings of the simulation models can be found in SI Appendix E. We emphasize that the simulations are only for demonstration purposes and they are based on reasonable assumptions. With future data and modeling support, the simulation models may be adjusted.

\begin{figure*}
\centering
\begin{tabular}{ccc}
\includegraphics[width=0.3\linewidth]{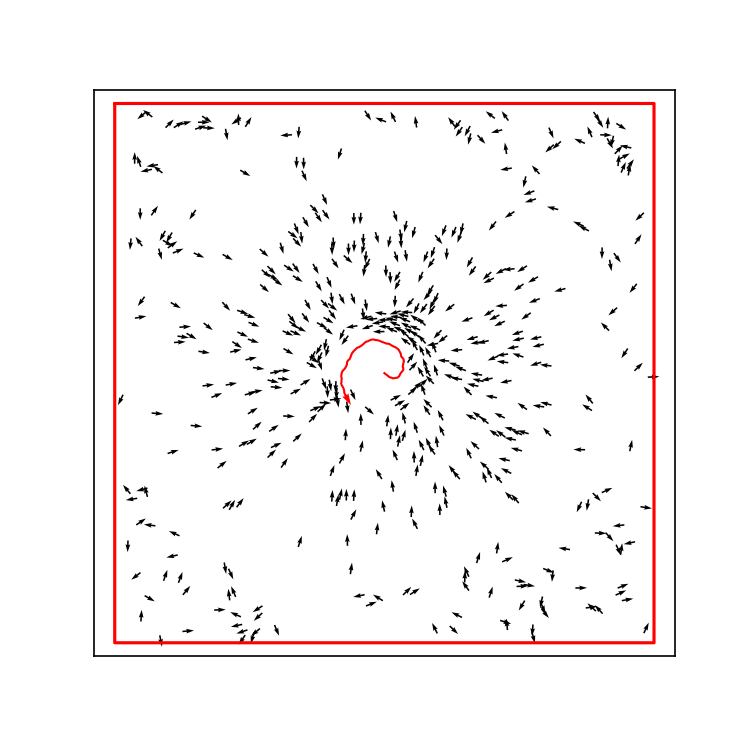}&
\includegraphics[width=0.3\linewidth]{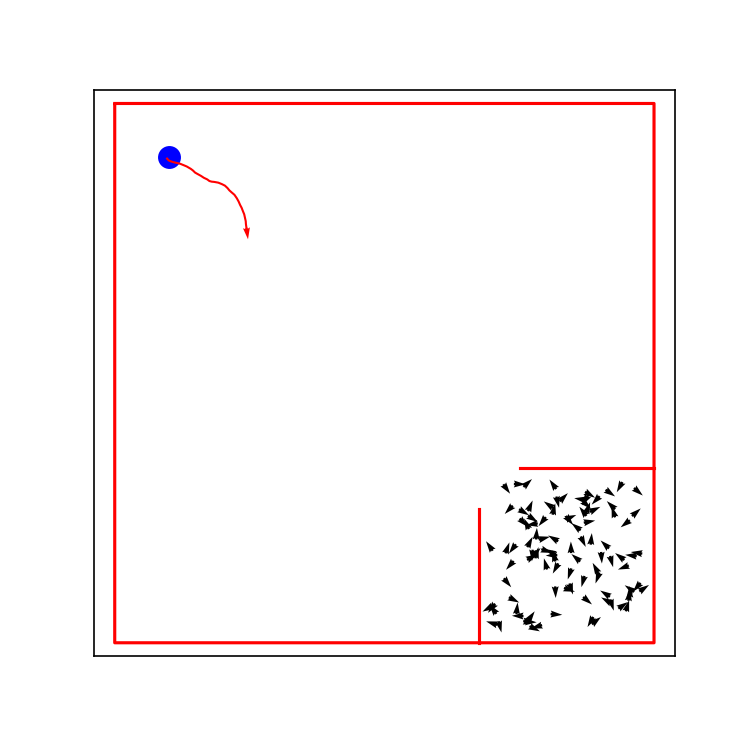}&
\includegraphics[width=0.3\linewidth]{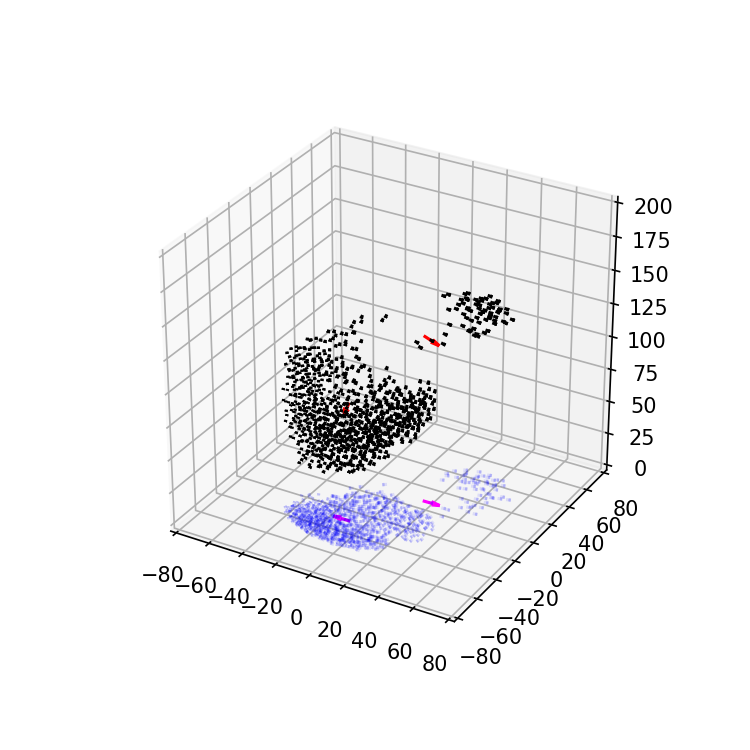}\\
A & C & E\\
\includegraphics[width=0.3\linewidth]{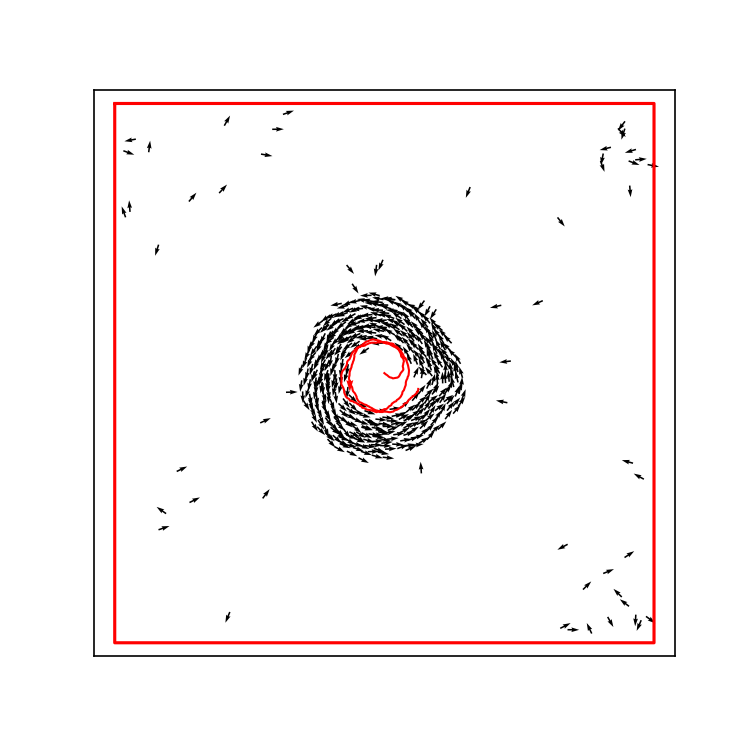}&
\includegraphics[width=0.3\linewidth]{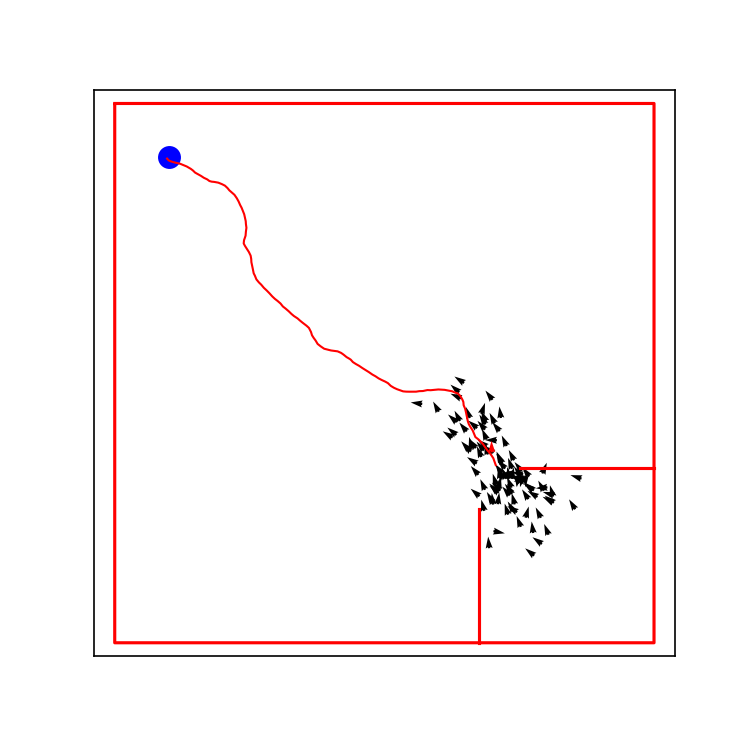}&
\includegraphics[width=0.3\linewidth]{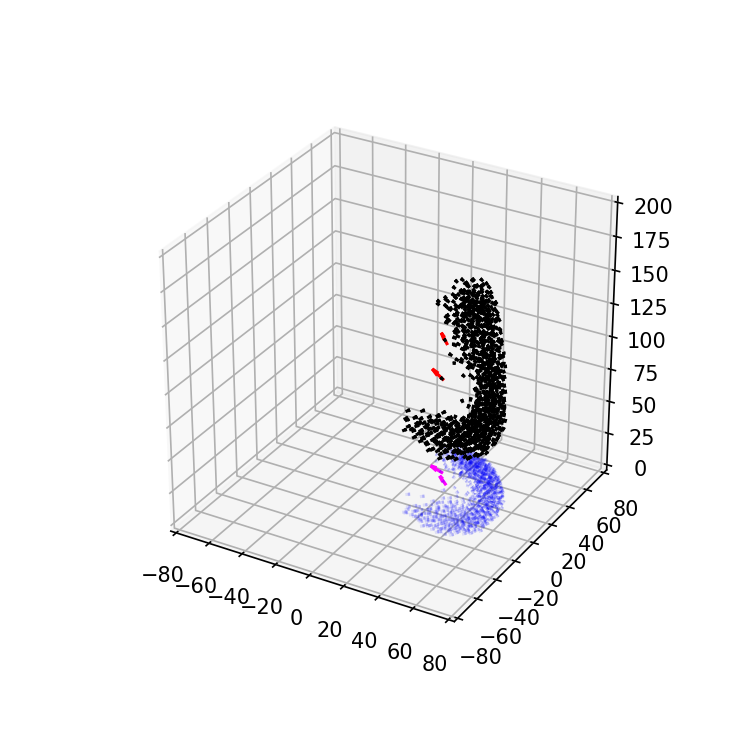}\\
B & D & F\\
\end{tabular}
\caption{
The snapshots of early and late stages of the simulations for (A,B) death mill of 500 army ants; (C,D) mass raids of 100 army ants; (E, F) flocking and escaping of 1000 birds. The red lines in (A,B,C,D) are the chemical trails produced by the leader. The blue dot in (C,D) is the food. The blue and pink points in the x-y plane of (E,F) are the projections from 3D to 2D. }
\label{fig:F4}
\end{figure*}

\begin{itemize}
\item {\bf Army ant death mill:} The virtually blind foragers of the New
World army ants \textit{Eciton burchelli} sometimes form a chemical (pheromone) trails system with a massive scale \cite{Couzin2003AntMill}. When some of the chemical trails form a circle, this can sometimes cause the phenomenon known as a death mill: a group of army ants forming a continuously rotating circle.
To simulate this, we assume that there is no attraction effect between individual ants, and adopt the alignment and repulsion effects formats of termites, as in section \ref{subsec:termite}. In addition, we define a particular ant individual as the ``leader'', who has the desire to create a circular chemical trail, and all the other individuals are followers who have the desire to follow the trail. The video of the simulation can be found in SI Video 2, and two snapshots of the simulation are presented in FIG. \ref{fig:F4} (A,B).

\item {\bf Army ant mass raids:} Several experiments were performed in reference \cite{Chandrae2021AntRaid} to study the foraging and raiding behaviors of army ants. The results showed that pheromone trail played a critical role in these phenomena. To simulate the behaviors described in reference \cite{Chandrae2021AntRaid}, we use a setting similar to the one we used for the army ant death mill, and then simply adjust the desire of the ``leader''. The behavior of the leader has three stages. First, the leader searches for the food. Second, once the leader finds the food, it returns to the nest and creates a pheromone trail for recruitment. Third, once the leader reaches the nest, it goes back to help obtain the food. All the other individuals have two stages of behaviors: wandering randomly before the recruitment and following the trail as a response to the recruitment. The video of the simulation can be found in SI Video 3, and two snapshots of the simulation are presented in FIG. \ref{fig:F4} (C,D).

\item {\bf The flocking and escaping of birds:} The beauty of the flocking and escaping behaviors of birds \cite{Youtube2020} involves multiple complex topics that remain challenging to explore \cite{Cavagna2018Flock, Hemelrijk2011SocialForce, Zumaya2018longrange, Cristiani2021Birds}. In this section, we describe an attempt to simulate this phenomenon with several assumptions. First, we combine the definitions of the interaction neighbors in \cite{Ballerini2008birds, Hemelrijk2011SocialForce, Zumaya2018longrange}, and we define the following interaction neighbors for our model: For the short-range interaction, each bird interacts with the nearest $n$ birds (topological). Within these nearest $n$ birds, interactions are classified into repulsion and/or alignment based on the distance (metric). It is observed that separated flocks of birds tend to merge together. Thus we also define the long-range attraction effect based on the distance (metric). We consider the ``roost'' effect as in \cite{Hemelrijk2011SocialForce} and define the boundary effect to restrict the flocks to a particular region. Finally, we introduce two predators that chase the nearest bird, and we define the desire of the birds as wanting to escape from the predators. The 2.5$\times$speed-up video of the simulation can be found in SI Video 4, and two snapshots of the simulation are presented in FIG. \ref{fig:F4} (E,F).

\end{itemize}

\section{Discussion}

In this work, we derive a general statistical mechanical modeling framework based on the principle of maximum entropy, the assumption of the time-delayed reaction, and the effects of the alignment, repulsion, attraction, boundary, and desire. This framework is very flexible in that some commonly used models, such as the Vicsek model, the social force model, and some of their variants can be considered special cases, as shown in Section \ref{subsec:other}. The advantage of expressing the equivalent discrete dynamic model (\ref{eqn:r}, \ref{eqn:v}) as the maximum entropy probability distribution is that it provides a powerful statistical tool to bridge the gap between experimental data and theoretical models. The corresponding log likelihood function can be used as a measurement for model comparison and parameter selection, as shown in section \ref{subsec:termite}. In addition to the results presented above, we would like to discuss several relevant issues.

First, we emphasize the fact that the assumption of delayed reaction time is necessary. Without the time-delay effect, the static model fails to capture the dynamic or non-equilibrium characteristics of the natural phenomena.
Furthermore, the absence of this assumption may cause ``global simultaneous interaction'': one turning of one animal may be reacted to by the whole system simultaneously, regardless of the size of the system. Due to the connection of the neighbor networks, such global simultaneous interaction may exist even when the interaction is only defined locally (See SI Appendix F for details). On the other hand, introducing the delay-time to this framework significantly simplifies the mathematical difficulty of modeling, especially when dealing with the asymmetric problem raised by the topological interaction neighbors (See SI Appendix F for details).
We also emphasize that it is possible that different individuals have different delayed reaction times, and the same individual can have different delayed reaction times to different effects. We assume a constant $\tau$ as an ``average'' for simplification, but one can of course introduce a more complicated delayed time system. Comparisons of different selections can be performed using log-likelihood values.

Second, it should be pointed out that the maximum entropy model does not guarantee the representation of the exact realistic model. The model only considers the constraints selected by the researcher and is evaluated using the observed data in the specified event, as mentioned in reference \cite{Bialek2012StatMech}. This framework is highly affected by the data and the selection of constraints. Even if we select the constraints that match the mechanics settings, it is possible that there are other factors that have been ignored or are very difficult to be captured (for example, the effect of wind in birds flocking). Furthermore, as a dynamic model that is trying to describe the behavior of living animals, we should be careful about the meaning of the physical terms for the living animals. For example, $\lak$ plays the role of ``average mass'' in our framework. However, equation (\ref{var}) indicates that it also measures the randomness of the motion of the system. It is possible that $\lak$ may change with respect to time, unlike the physical mass. Another example is that we can set the delay time $\tau=0$ in theory, but this may not match the reality of living animals (See SI Appendix F).

Finally, our modeling approach is based on the observed data for the velocities and positions, two sets of data that are relatively easier to collect. Each individual is taken as a particle without consideration of the shape, volume, or other factors that may affect the collective motion. To deal with this problem, adjustment can be made by adding additional direction and/or constraint, as in reference \cite{Helbing2000Human}. Another limitation of this framework is that it only considers the dot product. Thus, mechanics based on the cross product, such as the spin model in reference \cite{Cavagna2015birds} for turning, cannot directly be described by this framework. A rough approach to solve this problem is to assign the direction of centripetal force of turning to the direction of the boundary and/or desire effect.

Bridging the gap between theory and reality is always the challenge of modeling. Our flexible framework provides a potential tool for comparing various model assumptions and parameter selections for the same event. The best model may still be undiscovered. However, we can try to approach the best model via comparison. This is the meaning of our work.

\subsection*{Acknowledgement}
Work in China was supported by the National Natural Science Foundation of China (Grant No. 31772515).

\subsection*{Author Contribution}
C.W., J.C. and X.C. designed research; J.Z. and C.W. performed field studies; J.Z. and J.C. performed data analysis; J.C. performed modeling, theoretical analysis, computational analysis, programming, and simulations. J.C., C.W., and X.C. wrote the paper.

\subsection*{Data and Code Availability}
The data, the computational codes, and the simulation codes that have been used for this work are available from the corresponding author upon request when appropriate.

\subsection*{Conflict of Interest}
The authors declare no conflict of interest.

\nocite{*}

\input{Main.bbl}


\newpage
\quad

\newpage

\begin{center}
\onecolumngrid
\textbf{\large A General Statistical Mechanics Framework for the Collective Motion of Animals}\\
\quad\\
Jiacheng Cai, Jianlong Zhang, Xuan Chen, Cai Wang\\
\quad\\
\textbf{\large Supplementary Information\\
\quad\\}
\end{center}

\twocolumngrid

\appendix

\renewcommand{\theequation}{\thesection\arabic{equation}}
\renewcommand{\thefigure}{\thesection\arabic{figure}}
\renewcommand{\thetable}{\thesection\arabic{table}}

\input{SI}

\end{document}

%% file: Main.bbl
\providecommand{\noopsort}[1]{}\providecommand{\singleletter}[1]{#1}%
%

%% file: SI.tex
\section{\label{SIsec:model} The Model}

Inspired by the general idea of reference \cite{Bialek2012StatMech}, we construct a statistical mechanics framework for the collective motion of animals in this section. We first present the details of the principle of maximum entropy. Based on this, we derive our framework with the considerations of the delay reaction and several mechanics terms.

In general, we are looking for a probability density function $P(\bm{x})$, such that:

\begin{itemize}
    \item[1.] The entropy 
    \begin{equation}
    H[P]=-\int P(\bm{x})\ln P(\bm{x}) d\bm{x} \label{SIeqn:C1}
    \end{equation}
    is maximized.
    \item[2.] For some measurements $f_1(\bm{x}),\dots, f_k(\bm{x})$, The expectation constraints
    \begin{equation}
    	\Ep{f_w(\bm{x})} = \Eobs{f_w(\bm{x})}, w=1,\dots, k, \label{SIeqn:C2}
    \end{equation}
    are satisfied. $\Eobs{f}$ is the average of $f$ based on the observed data and $\Ep{f}$ is the expectation under $P$:
    \begin{equation*}
    	\Ep{f(\bm{x})} = \int f(\bm{x})P(\bm{x})d\bm{x}. 
    \end{equation*}
    \item[3.] As a probability density function, it requires
    \begin{equation}
    	\int P(\bm{x}) d\bm{x}=1. \label{SIeqn:C3}
    \end{equation}
\end{itemize}

The above problem can be approached by using the method of the Lagrange multipliers as follows:

We define a functional
\begin{eqnarray}
    H[P;\Lambda] &=& H[P] - \lambda_0\left(\int P(\bm{x}) d\bm{x}-1\right) \nonumber\\
                 &-& \sum_{w=1}^k\lambda_w\left(\Ep{f_w(\bm{x})} - \Eobs{f_w(\bm{x})}\right),
\end{eqnarray}
where $\Lambda=(\lambda_0,\lambda_1,\dots,\lambda_k)$ are the corresponding Lagrange multipliers. The corresponding optimization problem is equivalent to solving:
\begin{eqnarray}
    \frac{\partial H[P;\Lambda]}{\partial P(\bm{x})} &=& 0 \label{SIeqn:opt1}\\
    \frac{\partial H[P;\Lambda]}{\partial \lambda_w} &=& 0, \text{ for } w=1,\dots, k. \label{SIeqn:opt2}
\end{eqnarray}

Solving (\ref{SIeqn:opt1}) gives us
\begin{equation*}
    -\ln P(\bm{X})-1 - \lambda_0 - \sum_{w=1}^k\lambda_w f_w(\bm{x}) = 0.
\end{equation*}
Thus, 
\begin{equation}
    P(\bm{X}) = \frac{1}{Z(\Lambda)}\exp\left[ - \sum_{w=1}^k\lambda_w f_w(\bm{x})\right], \label{SIeqn:P0}
\end{equation}
where
\begin{equation}
    Z(\Lambda) = \exp(1+\lambda_0) = \int\exp\left[ - \sum_{w=1}^k\lambda_w f_w(\bm{x})\right]d\bm{x} \label{SIeqn:Zt}
\end{equation}
is the normalization factor to satisfy the condition (\ref{SIeqn:C3}). One can verify that the result distribution (\ref{SIeqn:P1}) satisfies the condition (\ref{SIeqn:C1}). 

Solving (\ref{SIeqn:opt2}) gives us
\begin{equation*}
    \Ep{f_w(\bm{x})} - \Eobs{f_w(\bm{x})} = 0.
\end{equation*}
Thus, 
\begin{equation}
    \frac{1}{Z(\Lambda)}\int f_w(\bm{x})\exp\left[ - \sum_{w=1}^k\lambda_w f_w(\bm{x})\right]d\bm{x}=\Eobs{f_w(\bm{x})},
\end{equation}
which satisfies the condition (\ref{SIeqn:C2}). In addition, the estimation of $\Lambda$ under this approach is equivalent to the maximum likelihood estimation, as shown in reference \cite{Bialek2012StatMech}.

Based on the above maximum entropy settings, the critical part of our modeling is finding the appropriate measurements $\bm{x}$ and $f_1(\bm{x}),\dots, f_k(\bm{x})$. Our goal is to construct a statistical mechanics framework to describe the collective motion of animals. We consider a group of $N$ animals whose motion state at time $t$ can be described by a set of vectors $\st=(\rt, \vt)$, where $\rt=(\rti{1},\rti{2},\cdots, \rti{N})$  and $\vt=(\vti{1},\vti{2},\cdots, \vti{N})$ are the positions and velocities of the animals at time $t$, respectively. $\bm{r}^t_i$ and $\bm{v}^t_i$ can be either two-dimensional ($d=2$) or three-dimensional ($d=3$) vectors, depending on the study interest. The motion state of the $i-$th animal at time $t$ is denoted by $\sti{i}=(\rti{i},\vti{i})$. 

Consider a time point $t$, we assume that the positions at time $t$ can be determined by the state information of the previous time step. Thus, we focus on the velocities and set $\bm{x}=\vt$, where $\bm{x}$ is the variable in the maximum entropy setting. 

We begin with the microscopic description at the $i-$th animal. Considering the similarity between the maximum entropy distribution (\ref{SIeqn:P0}) and the Boltzmann distribution in statistical mechanics, we choose the $f_w$ terms that can be related to some ``effective energy''. Naturally, the first term we select for the $i-$th animal is the ``kinetic energy'' term: 
\begin{equation}
        \fki{i} = \frac{1}{2}\vti{i}\cdot\vti{i},  \label{SIeqn:fki}
\end{equation}
It should be pointed out that the ``mass'' term is not included in the equation. As mentioned in the main text, it is managed by the Lagrange multiplier $\lak$.

In addition to the ``kinetic energy'', we consider some factors that may affect $\vti{i}$. In general, we assume that the effect of a factor $w$ consists of two parts: the direction term $\bm{u}$ and the strength term $g$. We also assume that the reaction of $\vti{i}$ to a factor is not completed at the same time as the factor is perceived. Thus, we assume that there exists a delay time $\tau>0$, such that $\bm{u}$ and $g$ are functions of $\stau$, and the corresponding effect to $\vti{i}$ is described as 
\begin{equation}
        \Qwi{i}{w} = \gwi{i}{w}{t-\tau}\uwi{i}{w}{t-\tau},  \label{SIeqn:Qwi}
\end{equation}
where $\eta_w$ is(are) the parameter(s) in the function $g$. 
The reaction of $\vti{i}$ to $\Qwi{i}{w}$ is 
\begin{equation}
        \fwi{i}{w} = -\Qwi{i}{w}\cdot\vti{i},  \label{SIeqn:fwi}
\end{equation}
where ``$\cdot$'' is the dot product for two vectors. We put the negative sign in the equation for mechanics purpose.

We divide various effects into two categories: the interactions between animals within the system and the external/internal effects. 

For the interactions, we consider the alignment, repulsion, and attraction effects, which are given by:
\begin{eqnarray}
    \hspace{-5ex}\Qwi{i}{ali}&=&\sum_{j\in\nalii}\gwi{i,j}{ali}{t-\tau}\vtaui{j},\\
    \hspace{-5ex}\Qwi{i}{rep}&=&\sum_{j\in\nrepi}\gwi{i,j}{rep}{t-\tau}\uwi{i,j}{rep}{t-\tau},\\
    \hspace{-5ex}\Qwi{i}{att}&=&\sum_{j\in\nattri}\gwi{i,j}{att}{t-\tau}\uwi{i,j}{att}{t-\tau},
\end{eqnarray}
where $\nalii, \nrepi, \nattri$ are the alignment, repulsion, and attraction neighbors, respectively, of the $i-$th animal at time $t-\tau$. If the $j-$th animal is within the corresponding neighbors at time $t-\tau$, then it has the corresponding effect to the $i-$th animal at time $t-\tau$, and this effect is reacted by the $i-$th animal at time $t$. It should be pointed out that the definitions of neighbors can be either metric \cite{Vicsek1995} or topological \cite{Ballerini2008} in our framework.
The direction terms are given by
\begin{eqnarray}
    \uwi{i,j}{rep}{t-\tau}&=&\frac{\rtaui{i}-\rtaui{j}}{\lVert\rtaui{i}-\rtaui{j}\rVert},\\
    \uwi{i,j}{att}{t-\tau}&=&\frac{\rtaui{j}-\rtaui{i}}{\lVert\rtaui{j}-\rtaui{i}\rVert}.
\end{eqnarray}

For the external/internal effects, we consider the boundary and desire effects, which are given by:
\begin{eqnarray}
    \Qwi{i}{bou} &=& \gwi{i}{bou}{t-\tau}\uwi{i}{bou}{t-\tau},\\
    \Qwi{i}{des} &=& \gwi{i}{des}{t-\tau}\uwi{i}{des}{t-\tau}.
\end{eqnarray}
Similarly, we can define $\nboi$ and $\nexi$ as the index set of the individuals that are affected by the ``boundary'' and ``desire'' effect, respectively, at time $t-\tau$ and respond at time $t$.
$\uboi$ is the ``boundary'' unit vector with the direction from the closest boundary point to $\rtaui{i}$, and $\uexti$ is the ``desire'' unit vector representing the direction of the will of the $i-$th animal. $\uboi$ and/or $\uexti$ terms are set to zero vectors when $i$ is not within the corresponding neighbor terms.

Up to this point, we have considered the ``kinetic energy'' and five factors that affect $\vti{i}$ from time $t-\tau$: the alignment, repulsion, and attraction interactions, and the boundary and desire effects. Other factors can be added via the same principle.
By extending the concept to all the animals in the group and taking the average, we obtain the following six measurements
\begin{eqnarray}
    \hspace{-3ex}\ft{kin} &=& \frac{1}{N}\sum_{i=1}^N\fki{i}, \\
    \hspace{-3ex}\ft{ali} &=& \frac{1}{N}\sum_{i=1}^N\falii{i}, \label{SIeqn:fks0}\\
    \hspace{-3ex}\ft{rep} &=& \frac{1}{N}\sum_{i=1}^N\frepi{i},\\
    \hspace{-3ex}\ft{att} &=& \frac{1}{N}\sum_{i=1}^N\fattri{i},\\
    \hspace{-3ex}\ft{bou} &=& \frac{1}{N}\sum_{i=1}^N\fboi{i},\\
    \hspace{-3ex}\ft{des} &=& \frac{1}{N}\sum_{i=1}^N\fexi{i}, \label{SIeqn:fdes}
\end{eqnarray}
with the corresponding Lagrange multipliers $\LAt=\left(\lak,\lali,\larep,\laattr,\labo,\laex\right)$. These are the average reactions of $\vt$ at time $t$ to the effects from time $t-\tau$.

To simplify the expression, We define the ``dot product'' for the data structures like $\vt$: let $\mathbf{a}=(\bm{a}_1,\dots,\bm{a}_N)$ and $\mathbf{b}=(\bm{b}_1,\dots,\bm{b}_N)$ where $\bm{a}_i=(a_{i1},\dots,a_{id}), \bm{b}_i=(b_{i1},\dots,b_{id})$ are d-dimensional vectors. The ``dot product'' of $\mathbf{a}$ and $\mathbf{b}$ is given by

\begin{equation}
\mathbf{a}\cdot\mathbf{b}=\sum_{i=1}^N\bm{a}_i\cdot\bm{b}_i=\sum_{i=1}^N\sum_{j=1}^d a_{ij}b_{ij}
\end{equation}

Substitute all the above $f$ terms and $\lambda$ terms to (\ref{SIeqn:P0}), we obtain
\begin{equation}
\Pt = \frac{1}{\Zt}\exp\left[\frac{1}{N}\left[-\frac{1}{2}\lak\vt\cdot\vt + \Ft\cdot\vt\right]\right],
\end{equation}
where $\Ft$ is an $N-$dimensional ``vector'' whose $i-$th component is a $d-$dimensional vector given by
\begin{eqnarray}
\Fti&=&\sum_{w}\lw{w}\Qwi{i}{w}\nonumber\\
&=&\lali\sum_{j\in\nalii}\gwi{i,j}{ali}{t-\tau}\vtaui{j} \nonumber\\
&+&\larep\sum_{j\in\nrepi}\gwi{i,j}{rep}{t-\tau}\uwi{i,j}{rep}{t-\tau} \nonumber\\
&+&\laattr\sum_{j\in\nattri}\gwi{i,j}{att}{t-\tau}\uwi{i,j}{att}{t-\tau} \nonumber\\
&+&\labo\gwi{i}{bou}{t-\tau}\uwi{i}{bou}{t-\tau} \nonumber\\
&+&\laex\gwi{i}{des}{t-\tau}\uwi{i}{des}{t-\tau} \label{SIeqn:Fti}
\end{eqnarray}
as the total effect from $\stau$ to $\vti{i}$.

Note that the $1/N$ term can be absorbed by $\Zt$, without the loss of the generality, we directly write
\begin{equation}
\Pt = \frac{1}{\Zt}\exp\left[-\frac{1}{2}\lak\vt\cdot\vt + \Ft\cdot\vt\right], \label{SIeqn:P1}
\end{equation}
where
\begin{eqnarray}
    \Zt &= &\int\exp\left[-\frac{1}{2}\lak\vt\cdot\vt + \Ft\cdot\vt\right]d\vt \nonumber\\  
        &= &\sqrt{\left(\frac{2\pi}{\lak}\right)^{dN}}\exp\left[\frac{1}{2\lak}\Ft\cdot\Ft\right] \nonumber\\
        \label{SIeqn:Zt2}
\end{eqnarray}
as a Gaussian integral. This is the main text equation (\ref{eqn:P0}). As mentioned in the main text, (\ref{SIeqn:P1}) is a multivariate normal distribution, the mean and covariance of which are given by
\begin{eqnarray}
\bm{\mu_{t,\tau}} &=& \frac{1}{\lak}\Ft, \label{SIeqn:mu}\\
\bm{\Sigma_{t,\tau}} &=& \frac{1}{\lak}\mathbf{I} \label{SIeqn:var},
\end{eqnarray}
where $\mathbf{I}$ is the $dN-$dimensional identity matrix.

The remain challenge is to estimate $\Eobs{\ft{w}}$ from the real-world data. By the meaning of $\Eobs{\ft{w}}$, one may need to set up the same $\stau$ and repeat the experiments multiple times, which can be very difficult or impossible for the studies of natural phenomena. Fortunately, since we define $\ft{w}=1/N \sum_{i=1}^N\fwi{i}{w}$ as the average reaction of $\vt$ to the corresponding effect, by the law of large numbers, we have
\begin{equation}
    \Eobs{\ft{w}}\approx f_{w}(\mathbf{v}_t^{obs}|\mathbf{s}_{t-\tau})
\end{equation} 
for large $N$ value, where $\mathbf{v}_t^{obs}$ is the observed data from one sample.

\section{\label{SIsec:others} Relationships with Other Models}

First, this framework can be specified as the Vicsek model and the social force model. For the $i-$th animal, the distribution (\ref{SIeqn:P1}), along with (\ref{SIeqn:mu},\ref{SIeqn:var}), can be rewritten as
\begin{eqnarray}
\rti{i}-\rtaui{i}&=&  \vtaui{i}\tau, \label{SIeqn:r}\\
\vti{i}-\vtaui{i}&=& (\Fti/\lak-\vtaui{i})+\bm{\epsilon_i}^{t,\tau}, \label{SIeqn:v}
\end{eqnarray}
where $\bm{\epsilon}_i^{t,\tau}$ is the white noise term that follows a d-dimensional normal distribution with the mean of $\bm{0}$ and the covariance of $1/\lak\mathbf{I}$. By this setting, $(\Fti-\lak\vtaui{i})/\tau$ is mathematically equivalent to the discrete approximation of the interaction forces and the forces from the boundary and desire effects. Since the general framework only specifies the directions of the effects from $\stau$ to $\vt$ and the detailed formats of the effects are not specified yet in $g$ terms, this framework has a large flexibility to be adjusted. Many widely used models can be considered special cases of this framework. We present the mapping between those models and our framework in the TABLE \ref{SItab:model}.

\begin{table*}
\caption{\label{SItab:model} Relationships of (\ref{SIeqn:v}) with other models}
\begin{ruledtabular}
\begin{tabular}{llll}
           & Law of inertia & Vicsek model \cite{Vicsek1995}                                            & Social force model without physical contact \cite{Helbing2000Human} \footnote{The meaning of $A,B,r_{ij},d_{ij},r_{i},d_{iW}$ can be found in reference \cite{Helbing2000Human}. $\lak,v_{i,des}$ denote $m, v_i^0$ in \cite{Helbing2000Human}, respectively. The physical contact forces among individuals or between individual and wall are not considered in this work. Detailed formats of the physical contact forces can be found in \cite{Helbing2000Human}. They can be easily added to our framework. }\\
\colrule
$\lali$    & $\lak$         & $\lak$                                                    & 0\\
$g_{ali}$  & 1              & $1/\Nalii$                                                & 0\\
$\nalii$   & $\{i\}$        & $\{j|\text{  } \lVert\rtaui{i}-\rtaui{j}\rVert<r\}$       & N/A\\
$\larep$   & 0              & 0                                                         & $A\tau$\\
$g_{rep}$  & 0              & 0                                                         & $\exp[(r_{ij}-d_{ij})/B]$\\
$\nrepi$   & N/A            & N/A                                                       & $\{j|\text{  } j\neq i\}$\\
$g_{att}$  & 0              & 0                                                         & 0\\
$\nattri$  & N/A            & N/A                                                       & N/A\\
$\labo$    & 0              & 0                                                         & $A\tau$\\
$g_{bou}$  & 0              & 0                                                         & $\exp[(r_{i}-d_{iW})/B]$\\
$\laex$    & 0              & 0                                                         & $\lak$\\
$g_{des}$  & 0              & 0                                                         & $v_{i,des}$\\
\end{tabular}

\end{ruledtabular}
\end{table*}

It addition, this framework can be extended to the maximum caliber model. The distribution (\ref{SIeqn:P1}) can be considered the probability of transition from $\stau$ to $\st$. Then, the motion states $(\stt{0},\stt{1},\dots,\stt{M})$ is a Markov chain, where $t_i=t_0+i\tau$. The joint distribution is given by

\begin{eqnarray}
    &\hspace{-6ex}\displaystyle P(\vtt{0},\vtt{1},\dots,\vtt{M})=P(\mathbf{v}_{t_0})\prod_{i=0}^{M-1} \Pti{i}, \nonumber\\
    &\displaystyle=\frac{P(\mathbf{v}_{t_0})}{\prod_{i=0}^{M-1}\Zti{i}}\exp\left[-\sum_{i=0}^{M-1}\sum_{w=1}^k \lambda_w^{t_i,\tau}f_w(\mathbf{v}_{t_{i+1}}|\mathbf{s}_{t_i})\right].\nonumber\\
    &\label{SIeqn:Pc}
\end{eqnarray}

With the assumption of $\lambda_w^{t_i,\tau}$ being constant for all $t_i$, the distribution (\ref{SIeqn:Pc}) is mathematically equivalent to the maximum caliber model with $\sum_{i=0}^{M-1} f_w(\mathbf{v}_{t_{i+1}}|\mathbf{s}_{t_i})$ being the constraints over the trajectories. A detailed survey of the principle of maximum caliber can be found in reference \cite{Presse2013MaxCal}.

\section{\label{SIsec:estimate} Parameter Estimations}

Given the $g$ and $\eta$ terms, as well as the data for $\st$ and $\stau$, we can obtain the maximum likelihood estimation for $\LAt$ from the distribution (\ref{SIeqn:P0}). In this section, we use $f,\mathbf{f}$ and $\mathbf{Q}$ terms as the quantities computed from the observational data without the notation ``obs'' for simplification.

Before the derivation, we first simplify the notations as follows:

\begin{itemize}
    \item In this section, we use $w$ as the general index of the name of the effects, such as ``$ali$'' or ``$bou$''. However, we do not use $w$ for ``$kin$'' due to its particularity. To consider the general case, we assume that there are $k$ effects exist in the system in addition to ``$kin$'' and label them as $w_1,\dots,w_k$. 
    
    \item We define $\fww=N\fw$ in (\ref{SIeqn:fks0}-\ref{SIeqn:fdes}). Similarly, we rewrite $f_{kin}^{t,\tau}=Nf_{kin}(\mathbf{v}_t|\mathbf{s}_{t-\tau})$.
    
\end{itemize}

With the above simplification, the distribution (\ref{SIeqn:P1}) becomes
\begin{equation}
    \Pt = \frac{1}{\Zt}\exp\left[-\lak\fwk-\sum_w\lw{w}\fww\right], \label{SIeqn:P3}
\end{equation}
where
\begin{align}
    &\Zt =\nonumber\\
    &\left(\frac{2\pi}{\lak}\right)^{\frac{dN}{2}}\exp\left[\frac{1}{2\lak}\sum_{u,v}\lw{w_u}\lw{w_v}\sum_{i=1}^N\Qwi{i}{w_u}\cdot\Qwi{i}{w_v}\right]. \label{SIeqn:Z3}
\end{align}

The corresponding log-likelihood function is
\begin{equation}
    \ln P = -\ln\Zt -\lak\fwk-\sum_w\lw{w}\fww, \label{SIeqn:lnP}
\end{equation}
where
\begin{eqnarray}
  \hspace{-3ex}  \ln\Zt &=& \frac{dN}{2}\ln(2\pi)-\frac{dN}{2}\ln(\lak)\nonumber\\
    &+& \frac{1}{2\lak}\sum_{u,v}\lw{w_u}\lw{w_v}\sum_{i=1}^N\Qwi{i}{w_u}\cdot\Qwi{i}{w_v}.\label{SIeqn:lnZ}
\end{eqnarray}

(\ref{SIeqn:lnP}) is maximized by solving the following system of equations for $\LAt=\{\lak,\lw{w_1},\lw{w_2},\dots,\lw{w_k}\}$:
\begin{equation}
    -\frac{\partial}{\partial\lak}\ln\Zt-\fwk=0,
\end{equation}
which gives us
\begin{equation}
    \frac{dN}{2}\frac{1}{\lak}+\frac{1}{2}\sum_{u,v}\frac{\lw{w_u}}{\lak}\frac{\lw{w_v}}{\lak}\sum_{i=1}^N\Qwi{i}{w_u}\cdot\Qwi{i}{w_v}=\fwk. \label{SIeqn:lk}
\end{equation}
And
\begin{equation}
    -\frac{\partial}{\partial\lw{w_u}}\ln\Zt-\fwu{w_u}=0, \text{ } u=1,\dots, k,
\end{equation}
which gives us
\begin{equation}
    \frac{\lw{w_u}}{\lak}\sum_{i=1}^N\Qwi{i}{w_u}\cdot\Qwi{i}{w_u}+\sum_{v\neq u}\frac{\lw{w_v}}{\lak}\sum_{i=1}^N\Qwi{i}{w_u}\cdot\Qwi{i}{w_v}=-\fwu{w_u}. \label{SIeqn:lw}
\end{equation}

Let $\xi_u^{t,\tau}=\lw{w_u}/\lak$. The equations (\ref{SIeqn:lw}) can be written as
\begin{equation}
\mathbf{Q}\bm{\xi} = \mathbf{f},\label{SIeqn:lw00}
\end{equation}
where $\mathbf{Q}$ is a $k\times k$ matrix whose $u,v-$th element is $\sum_{i=1}^N\Qwi{i}{w_u}\cdot\Qwi{i}{w_v}$; $\bm{\xi}=(\xi_1^{t,\tau}, \dots, \xi_k^{t,\tau})^T$; $\mathbf{f}=(-\fwu{w_1}, \dots, -\fwu{w_k})^T$, and $T$ represents the transpose operation. If $\mathbf{Q}$ is invertible, then we can obtain the maximum likelihood estimators for $\bm{\xi}$ as
\begin{equation}
\bm{\xi}^* = \mathbf{Q}^{-1}\mathbf{f}. \label{SIeqn:lw0}
\end{equation}

The equation (\ref{SIeqn:lk}) can be rewritten as
\begin{equation}
    \frac{dN}{2}\frac{1}{\lak}+\frac{1}{2}(\bm{\xi}^*)^T\mathbf{Q}\bm{\xi}^*=\fwk. \label{SIeqn:lk2}
\end{equation}

Substituting (\ref{SIeqn:lw00}, \ref{SIeqn:lw0}) to (\ref{SIeqn:lk2}) and considering that $\mathbf{Q}$ is symmetric, we obtain the maximum likelihood estimator for $\lak$ as
\begin{equation}
\laks = \frac{dN}{2\fwk-\mathbf{f}^T\mathbf{Q}^{-1}\mathbf{f}}. \label{SIeqn:lk0}
\end{equation}

One may understand the estimation formula (\ref{SIeqn:lk0}) as follows: $2\fwk/(dN)=\vt\cdot\vt/(dN)$ is a measurement of the average observed squared speed over the group and dimensions, while $\mathbf{f}^T\mathbf{Q}^{-1}\mathbf{f}/(dN)$ can be considered the similar estimation with all the effects of the factors removed (consider the formulas and meanings of $\mathbf{f}$ and $\mathbf{Q}$). When the difference between these two quantities is small, it means that, on average, the system is not affected by the factors significantly,  which corresponds to a larger $\laks$ value.

In summary, the maximum likelihood estimators for $\LAt$ is given by
\begin{equation}
\LAtss = (\laks, \laks\mathbf{f}^T\mathbf{Q}^{-1}). \label{SIeqn:LAt0}
\end{equation}

It should be pointed out that some particular models may require two or more $\lambda$ terms equal to each other. For example, the Vicsek model requires $\lali=\lak$, while the Social force model requires $\larep=\labo$ and $\laex=\lak$ (See TABLE \ref{SItab:model}). We discuss the adjustments of the above algorithm for the following two cases:

\begin{itemize}
    \item When there exist $u\neq v$ such that $\lw{w_u}=\lw{w_v}$, we merge the $w_v$ term into the $w_u$ term. The equation (\ref{SIeqn:lw}) becomes:
    
    \begin{widetext}
    \begin{equation}
    -\fwu{w_u}-\fwu{w_v} = \sum_{j\neq u,v}\frac{\lw{w_j}}{\lak}\sum_{i=1}^N\left(\Qwi{i}{w_u}\cdot\Qwi{i}{w_j}+\Qwi{i}{w_v}\cdot\Qwi{i}{w_j}\right) 
    +  \frac{\lw{w_u}}{\lak}\sum_{i=1}^N\left(\Qwi{i}{w_u}\cdot\Qwi{i}{w_u}+\Qwi{i}{w_v}\cdot\Qwi{i}{w_v}\right).\label{SIeqn:lwcase1}
    \end{equation}
    \end{widetext}
    
    The dimensions of $\mathbf{Q}, \bm{\xi}$, and $\mathbf{f}$ are reduced by 1, and the corresponding elements should be adjusted based on (\ref{SIeqn:lwcase1}).
    
    \item When there exists $u$ such that $\lw{w_u}=\lak$, the equation (\ref{SIeqn:lk}) becomes
    
    \begin{equation}
    \frac{dN}{2}\frac{1}{\lak}+\frac{1}{2}\sum_{j,v\neq u}\frac{\lw{w_j}}{\lak}\frac{\lw{w_v}}{\lak}\sum_{i=1}^N\Qwi{i}{w_j}\cdot\Qwi{i}{w_v}=\fwk+\fwu{w_u}. \label{SIeqn:lkcase2}
    \end{equation}
    
    The maximum likelihood estimators become
    
    \begin{eqnarray}
        \laks &=& \frac{dN}{2\fwk+2\fwu{w_u}-\mathbf{f}_*^T\mathbf{Q}_*^{-1}\mathbf{f}_*}, \label{SIeqn:lk1}\\
        \LAtss &=& (\laks, \laks\mathbf{f}_*^T\mathbf{Q}_*^{-1}),\label{SIeqn:LAt1}
    \end{eqnarray}
    
    where $\mathbf{f}_*$ and $\mathbf{Q}_*$ are $\mathbf{f}$ and $\mathbf{Q}$ removing the $u-$th element/row and column, respectively. 
    
\end{itemize}

\section{\label{SIsec:experiment} The Experiment}

The description of the experiment can be found in the main text section I C. In this section, we present the detailed settings of the two models that are used as a demonstration of model comparisons via our framework.  

As mentioned in the main text, we only consider the alignment, repulsion, boundary, and desire effects during the short escape time period. The following settings are the same for the two models:
\begin{eqnarray*}
&\nalii  &= \{j|\text{ } \lVert\rtaui{i}-\rtaui{j}\rVert\leq l_0\};\\
&\nrepi  &= \{j|\text{ } 0<\lVert\rtaui{i}-\rtaui{j}\rVert\leq 0.5 l_0\};\\
&g_{rep} &= 1;\\
&\nboi   &= \{i|\text{ } 0<\lVert\rtaui{i}-\rboi\rVert\leq 0.5 l_0 \}; \\
&g_{bou} &= 1;\\
&\nexi   &= \{i|\text{ all $i$ that are within the boundary}  \}; \\
&g_{des} &= 1;\\
&\uexti  &= (0,-1).
\end{eqnarray*}

$l_0=0.389$ cm is the average body length of termites in this field study.
The above settings can be considered the simplification of the social force model. 

The factor that we are interested in is whether the format of alignment follows the Vicsek setting. We set $g_{ali}=1/\Nalii$ and require $\lali=\lak$ for the Vicsek setting, and set $g_{ali}=1$ without the restriction of $\lali$ for the non-Vicsek setting.

With the above $g$ and $n$ terms, as well as the state data $\st$ and $\stau$ for all combinations of $(t,\tau)$, the maximum likelihood estimations for $\LAt$ for the two models can be obtained via equations (\ref{SIeqn:lk0},\ref{SIeqn:LAt0}) and (\ref{SIeqn:lk1},\ref{SIeqn:LAt1}). Then, the corresponding log-likelihood values can be computed via equations (\ref{SIeqn:lnP},\ref{SIeqn:lnZ}). The results for these computations are shown in TABLE \ref{SItab:exp}. The model with the Vicsek setting has larger log-likelihood values than the non-Vicsek setting at all combinations of $(t,\tau)$. 
Finally, we compare the probability density histograms for the actual observed data and the simulations based on the model with the Vicsek setting. We use $\tau^*=0.033s$, as it has the maximum likelihood over all $t$. We present the results for $t=0.133s$ in the main text FIG. 3, and present the results for $t=4.133, 9.133s$ in FIG. \ref{SIfig:SIF1},\ref{SIfig:SIF2} , respectively.

\setcounter{figure}{0}
\setcounter{table}{0}

\begin{table*}
\caption{\label{SItab:exp} Summary of two models for describing the emergent behavior of termites}
\begin{ruledtabular}
\begin{tabular}{lll|rrrrr|rrrrrr}
  &   &        & \multicolumn{5}{c|}{Vicsek} & \multicolumn{6}{c}{non-Vicsek}\\
t & N & $\tau$ & $\ln P$ & $\lak$  & $\larep$ & $\labo$ & $\laex$ & $\ln P$ & $\lak$ & $\lali$ & $\larep$ & $\labo$ & $\laex$\\
\colrule
0.133 &384 &0.033 &-527.433 &4.119 &0.200 &-0.417 &0.083 &-613.567 &3.456 &0.187 &0.902 &0.015 &0.478 \\
~ &~ &0.067 &-594.895 &3.415 &0.202 &-0.252 &0.101 &-652.142 &3.125 &0.204 &0.716 &-0.067 &0.617 \\
~ &~ &0.100 &-638.119 &2.807 &0.193 &-0.052 &0.242 &-672.067 &2.967 &0.207 &0.638 &0.071 &0.768 \\
~ &~ &0.133 &-645.636 &2.657 &0.114 &-0.017 &0.306 &-677.667 &2.924 &0.142 &0.630 &0.007 &0.803 \\
\colrule
1.133 &384 &0.033 &-460.377 &4.767 &0.228 &0.341 &0.099 &-598.817 &3.591 &0.219 &0.748 &0.194 &0.696 \\
~ &~ &0.067 &-560.103 &3.273 &0.208 &-0.008 &0.246 &-626.310 &3.343 &0.230 &0.694 &0.038 &0.828 \\
~ &~ &0.100 &-627.075 &2.770 &0.073 &0.261 &0.243 &-658.783 &3.072 &0.076 &0.570 &0.125 &0.955 \\
~ &~ &0.133 &-651.777 &2.582 &0.082 &0.013 &0.248 &-676.603 &2.933 &0.105 &0.507 &-0.023 &0.990 \\
\colrule
2.133 &377 &0.033 &-362.014 &4.928 &0.075 &-0.034 &0.306 &-532.078 &4.164 &0.073 &0.677 &-0.040 &1.625 \\
~ &~ &0.067 &-462.903 &3.571 &0.098 &-0.133 &0.362 &-555.856 &3.910 &0.138 &0.604 &-0.282 &1.748 \\
~ &~ &0.100 &-491.455 &3.178 &0.169 &0.016 &0.408 &-564.952 &3.816 &0.180 &0.541 &-0.168 &1.863 \\
~ &~ &0.133 &-518.054 &2.772 &0.137 &0.178 &0.481 &-576.484 &3.702 &0.143 &0.498 &0.104 &1.956 \\
\colrule
3.133 &329 &0.033 &-249.921 &6.830 &0.485 &0.439 &0.163 &-415.583 &4.829 &0.355 &0.778 &0.502 &1.790 \\
~ &~ &0.067 &-350.351 &4.732 &0.113 &0.296 &0.250 &-433.356 &4.575 &0.157 &0.724 &0.203 &1.832 \\
~ &~ &0.100 &-387.488 &3.956 &0.135 &0.299 &0.320 &-434.451 &4.560 &0.224 &0.783 &0.336 &1.794 \\
~ &~ &0.133 &-397.175 &3.437 &0.224 &0.208 &0.448 &-445.564 &4.409 &0.294 &0.689 &0.171 &1.986 \\
\colrule
4.133 &327 &0.033 &-270.857 &5.137 &0.143 &0.348 &0.410 &-390.516 &5.174 &0.135 &0.884 &0.525 &1.740 \\
~ &~ &0.067 &-351.482 &4.262 &0.125 &0.124 &0.346 &-414.464 &4.809 &0.117 &0.793 &0.431 &1.777 \\
~ &~ &0.100 &-358.127 &4.148 &0.159 &0.111 &0.351 &-417.518 &4.764 &0.227 &0.780 &0.488 &1.794 \\
~ &~ &0.133 &-387.752 &3.357 &0.178 &0.143 &0.484 &-419.644 &4.733 &0.302 &0.787 &0.506 &1.900 \\
\colrule
5.133 &309 &0.033 &-198.470 &8.395 &0.478 &-0.135 &0.071 &-332.371 &5.825 &0.373 &0.962 &-0.133 &1.736 \\
~ &~ &0.067 &-281.466 &6.466 &0.197 &-0.274 &0.070 &-352.482 &5.458 &0.205 &0.832 &-0.215 &1.834 \\
~ &~ &0.100 &-295.293 &5.716 &0.108 &-0.632 &0.154 &-371.918 &5.126 &0.122 &0.735 &-0.546 &1.923 \\
~ &~ &0.133 &-317.377 &4.691 &0.065 &-0.389 &0.310 &-371.976 &5.125 &0.148 &0.758 &-0.356 &1.928 \\
\colrule
6.133 &269 &0.033 &-157.863 &8.506 &0.121 &-0.053 &0.131 &-280.285 &6.025 &0.145 &1.014 &0.098 &1.706 \\
~ &~ &0.067 &-243.545 &5.467 &-0.049 &-0.307 &0.276 &-301.441 &5.569 &0.068 &0.901 &-0.338 &1.838 \\
~ &~ &0.100 &-267.715 &4.451 &0.002 &-0.516 &0.414 &-299.340 &5.613 &0.065 &0.927 &-0.532 &1.985 \\
~ &~ &0.133 &-285.586 &3.788 &-0.037 &-0.331 &0.535 &-302.652 &5.544 &0.003 &0.907 &-0.335 &2.091 \\
\colrule
7.133 &245 &0.033 &-82.533 &11.192 &0.747 &-0.321 &0.081 &-252.125 &6.103 &0.450 &0.731 &-0.204 &2.581 \\
~ &~ &0.067 &-162.765 &7.681 &0.437 &-0.281 &0.147 &-267.194 &5.739 &0.426 &0.641 &-0.242 &2.687 \\
~ &~ &0.100 &-204.833 &4.882 &0.230 &-0.173 &0.487 &-263.990 &5.815 &0.340 &0.736 &0.074 &2.672 \\
~ &~ &0.133 &-208.865 &4.743 &0.393 &-0.181 &0.493 &-264.740 &5.797 &0.494 &0.720 &0.128 &2.711 \\
\colrule
8.133 &222 &0.033 &-82.785 &10.615 &0.487 &0.503 &0.117 &-224.989 &6.199 &0.376 &0.732 &0.675 &2.299 \\
~ &~ &0.067 &-135.229 &8.083 &0.407 &1.161 &0.145 &-241.294 &5.760 &0.294 &0.597 &0.907 &2.437 \\
~ &~ &0.100 &-133.902 &7.623 &0.514 &0.887 &0.234 &-233.666 &5.962 &0.488 &0.694 &0.820 &2.380 \\
~ &~ &0.133 &-165.655 &6.507 &0.280 &0.748 &0.266 &-242.864 &5.720 &0.233 &0.607 &0.755 &2.429 \\
\colrule
9.133 &184 &0.033 &-64.804 &11.460 &0.247 &-0.021 &0.066 &-166.674 &6.904 &0.052 &1.240 &-0.216 &1.875 \\
~ &~ &0.067 &-133.753 &7.824 &0.444 &-0.457 &0.063 &-185.249 &6.241 &0.275 &0.979 &-0.266 &1.997 \\
~ &~ &0.100 &-134.353 &7.484 &0.190 &-0.727 &0.117 &-190.207 &6.075 &0.103 &0.879 &-0.635 &2.076 \\
~ &~ &0.133 &-144.735 &6.773 &0.705 &0.080 &0.176 &-191.812 &6.022 &0.561 &0.886 &0.006 &2.074 \\

\end{tabular}
\end{ruledtabular}
\end{table*}

\begin{figure*}
\centering
\begin{tabular}{cc}
\includegraphics[width=0.4\linewidth]{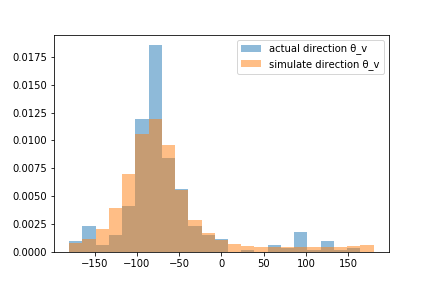}&
\includegraphics[width=0.4\linewidth]{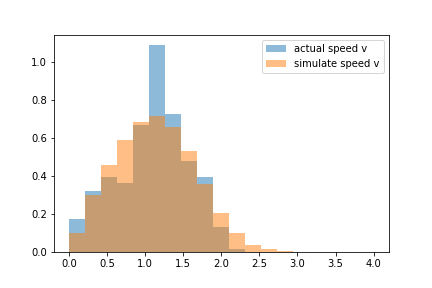}\\
A & B\\
\includegraphics[width=0.4\linewidth]{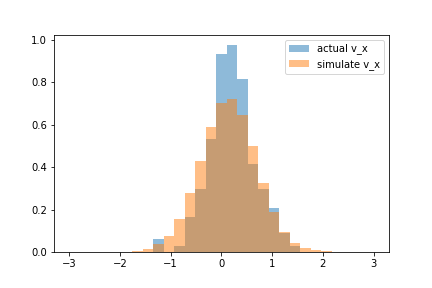}&
\includegraphics[width=0.4\linewidth]{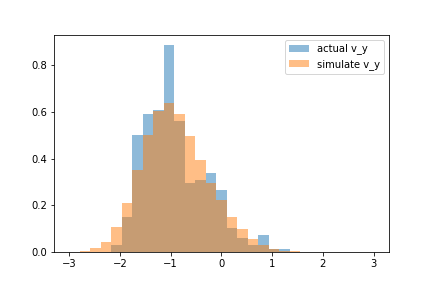}\\
C & D\\
\end{tabular}
\caption{
Comparisons of probability density histograms between the actual observed data and the 1000 simulations for all termites (at $t=4.133s$ with $\tau^*=0.033s$). Simulations are based on the model with the Vicsek setting. Figures show the comparisons for : (A) velocity angles in the unit of degree; (B) speed in the unit of cm/s; (C) projection of velocities on the $x$ direction in the unit of cm/s; (D) projection of velocities on the $y$ direction in the unit of cm/s.}
\label{SIfig:SIF1}
\end{figure*}

\begin{figure*}
\centering
\begin{tabular}{cc}
\includegraphics[width=0.4\linewidth]{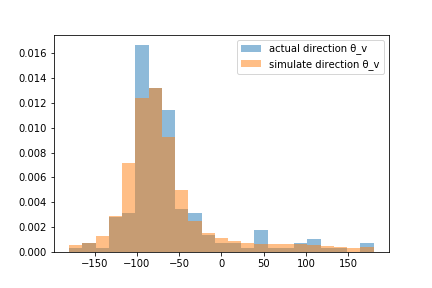}&
\includegraphics[width=0.4\linewidth]{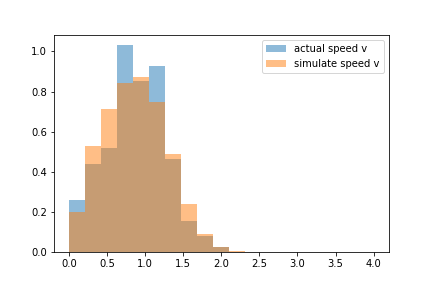}\\
A & B\\
\includegraphics[width=0.4\linewidth]{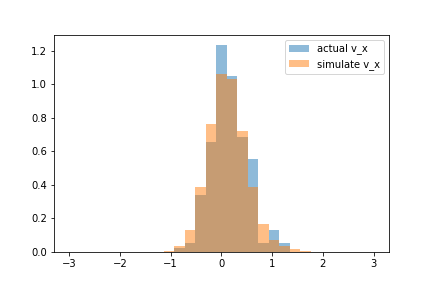}&
\includegraphics[width=0.4\linewidth]{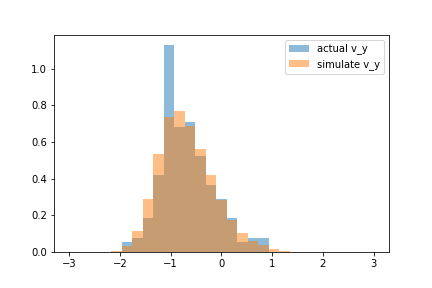}\\
C & D\\
\end{tabular}
\caption{
Comparisons of probability density histograms between the actual observed data and the 1000 simulations for all termites (at $t=9.133s$ with $\tau^*=0.033s$). Simulations are based on the model with the Vicsek setting. Figures show the comparisons for : (A) velocity angles in the unit of degree; (B) speed in the unit of cm/s; (C) projection of velocities on the $x$ direction in the unit of cm/s; (D) projection of velocities on the $y$ direction in the unit of cm/s.}
\label{SIfig:SIF2}
\end{figure*}

\section{\label{SIsec:simulation} The Simulations}

The descriptions of the three simulation experiments can be found in the main text section I D. In this section, we present the detailed settings of the $\lambda, g,n,\bm{u}$ terms, and other particular issues for the simulations. We emphasize that the simulations are just for demonstration purposes. The purpose of the setting selections is to reproduce the phenomena of collective motion observed in nature and labs. Thus, we do not consider the units in all three simulations, unless specified otherwise.

\begin{itemize}
    \item {\bf Army ant death mill:} $N$ agents are randomly placed in a $20\times20$ square region. They are assigned random initial velocities that identically and independently follow the 2-dimensional normal distributions with mean of zero and covariance of 0.01$\mathbf{I}$. We simulate $M$ time steps. We choose the following settings.
    \begin{eqnarray*}
    &l_0     &= 0.5;\\
    &N       &= 500;\\
    &M       &= 500;\\
    &\tau    &= 0.03s;\\
    &\lali   &= \lak;\\ 
    &\nalii  &= \{j|\text{ } \lVert\rtaui{i}-\rtaui{j}\rVert\leq 1.2 l_0\};\\
    &g^{t,\tau}_{i,ali}  &= 1/\Nalii;\\
    &\larep  &= 3.0;\\
    &\nrepi  &= \{j|\text{ } 0<\lVert\rtaui{i}-\rtaui{j}\rVert\leq l_0\};\\
    &g_{rep} &= 1.0;\\
    &\labo   &= 40.0;\\
    &\nboi   &= \{i|\text{ } 0<\lVert\rtaui{i}-\rboi\rVert\leq  l_0 \}; \\
    &g_{bou} &= 1.0;\\
    &\laex   &= 3.0;\\
    &g_{des} &= 1.0.
    \end{eqnarray*}
    
    As mentioned in the main text, we define the individual with the index of $0$ as the leader to create a chemical trail. We put the leader to the center of the region and assign the initial velocity as (1,0). The leader has $\lakle=2$, and its desire direction at the $k-$th time step $t=k\tau$ is defined by
    \begin{eqnarray*}
    \uaexle  &=& (\uaexlex,\uaexley),\\
    \uaexlex &=& \uaexlexs/\lVert(\uaexlexs,\uaexleys)\rVert,\\
    \uaexley &=& \uaexleys/\lVert(\uaexlexs,\uaexleys)\rVert,
    \end{eqnarray*}
    where
    \begin{eqnarray*}
    \uaexlexs &=& \cos(8k\pi/M)-8k\pi\sin(8k\pi/M)/M,\\
    \uaexleys &=& \sin(8k\pi/M)+8k\pi\cos(8k\pi/M)/M.
    \end{eqnarray*}
    This direction of motion encourages the leader to move along an Archimedean spiral. This setting can balance the repulsion force caused by the gathering ants to the leader, as shown in SI Video 2. The positions and directions of the points in the chemical trail up to the $k-$th time step are recorded as 
    \begin{eqnarray*}
    \mathbf{Tr}_k &=& (\bm{r}^{t_0}_0, \bm{r}^{t_1}_0, \cdots, \bm{r}^{t_k}_0),\\
    \mathbf{Di}_k &=& \left(\frac{\bm{v}^{t_0}_0}{\lVert\bm{v}^{t_0}_0\rVert}, \frac{\bm{v}^{t_1}_0}{\lVert\bm{v}^{t_1}_0\rVert}, \cdots, \frac{\bm{v}^{t_k}_0}{\lVert\bm{v}^{t_k}_0\rVert}\right).\\
    \end{eqnarray*}
    We assume that the leader only creates the trail in the first 300 steps.
    
    The rest individuals of the ant colony are the followers. To demonstrate the forming of the death mill, the ant colony is set not to notice the existence of the trajectory during the first 50 steps. During this period, we set $\lak=10$, and no desire effect is assigned to the followers. After the first 50 steps, we set  $\lak=\lakle$. The desire direction of the $i-$th ant at time $t=k\tau$ during this period is defined as
    
    \begin{equation*}
        \uexti=
        \begin{cases}
            \bm{0}, & \lVert\bm{r}^{t_j}_0-\rtaui{i}\rVert>30l_0, \\
            \frac{\bm{r}^{t_j}_0-\rtaui{i}}{\lVert\bm{r}^{t_j}_0-\rtaui\rVert}, & 6l_0<\lVert\bm{r}^{t_j}_0-\rtaui{i}\rVert\leq 30l_0, \\
            \frac{\bm{r}^{t_j}_0-\rtaui{i}}{2\lVert\bm{r}^{t_j}_0-\rtaui{i}\rVert} + \frac{\bm{v}^{t_j}_0}{\lVert\bm{v}^{t_j}_0\rVert}, & 2l_0<\lVert\bm{r}^{t_j}_0-\rtaui{i}\rVert\leq 6l_0, \\
            \frac{\bm{v}^{t_j}_0}{\lVert\bm{v}^{t_j}_0\rVert}, & \lVert\bm{r}^{t_j}_0-\rtaui{i}\rVert\leq 2l_0,
        \end{cases}
    \end{equation*}
    where $\bm{r}^{t_j}_0\in\mathbf{Tr}_k$ is the closest point of the trail to $\rtaui{i}$.
    
    Finally, we assume that there exists a speed limit to the ants. The maximum speed of the leader is 3. We set the maximum speed for the followers as 1 for the first 50 steps, then 3 when forming the mill.
    
    \item {\bf Army ant mass raids:} The settings of this simulation are basically the same as the settings in the army ant death mill, except for the following:
    
    First, a nest region is added in the bottom right corner, of which the exit position is $\bm{r}_n$. Food is added in the upper left corner at the position of $\bm{r}_f$. The leader is placed near the food and the followers are placed in the nest.
    
    Second, we adjust the following setting:
    \begin{eqnarray*}
    &l_0     &= 0.2;\\
    &N       &= 100;\\
    &\nalii  &= \{j|\text{ } \lVert\rtaui{i}-\rtaui{j}\rVert\leq l_0\};\\
    &\nrepi  &= \{j|\text{ } 0<\lVert\rtaui{i}-\rtaui{j}\rVert\leq 0.5l_0\};\\
    &\labo   &= 50.0;\\
    &\laex   &= 1.0.
    \end{eqnarray*}
    
    Third, we define the ``flag'' variable to indicate the stage of the behavior. For the leader, we have 
    
    \begin{equation*}
        \text{Flag}_l=
        \begin{cases}
            0 & \text{when searching for food}, \\
            1 & \text{food found and returning to nest}, \\
            2 & \text{reaches nest and returning to food.} 
        \end{cases}
    \end{equation*}
    
    For the followers, we have
    
    \begin{equation*}
        \text{Flag}_f=
        \begin{cases}
            0 & \text{when $\text{Flag}_l=0$ or 1}, \\
            1 & \text{when $\text{Flag}_l=2$}.
        \end{cases}
    \end{equation*}
     
     Then, the corresponding desire direction of the leader is given by
     
     \begin{equation*}
        \uaexle=
        \begin{cases}
            \frac{\bm{r}_f-\rtaui{0}}{\lVert\bm{r}_f-\rtaui{0}\rVert}, & \text{when $\text{Flag}_l=0$},\\
            \frac{\bm{r}_n-\rtaui{0}}{\lVert\bm{r}_n-\rtaui{0}\rVert}, & \text{when $\text{Flag}_l=1$},\\
            \text{same format as } \uexti, & \text{when $\text{Flag}_l=2$},
        \end{cases}
    \end{equation*}
    
    where $\uexti$ is the desire direction of the followers at $\text{Flag}_f=1$, defined below
    
    Similar to the death mill, we can define the chemical trail $\{\mathbf{Tr}, \mathbf{Di}\}$ that is created by the leader during the stage of $\text{Flag}_l=1$. 
    
    For the followers, $\uexti$ is set to zero when $\text{Flag}_f=0$. When $\text{Flag}_f=1$, we have
    
    \begin{equation*}
        \uexti=
        \begin{cases}
            \bm{0}, & \lVert\bm{r}^{t_j}_0-\rtaui{i}\rVert>80l_0, \\
            \frac{\bm{r}^{t_j}_0-\rtaui{i}}{\lVert\bm{r}^{t_j}_0-\rtaui\rVert}, & 20l_0<\lVert\bm{r}^{t_j}_0-\rtaui{i}\rVert\leq 80l_0, \\
            \frac{\bm{r}^{t_j}_0-\rtaui{i}}{2\lVert\bm{r}^{t_j}_0-\rtaui{i}\rVert} - \frac{\bm{v}^{t_j}_0}{\lVert\bm{v}^{t_j}_0\rVert}, & 10l_0<\lVert\bm{r}^{t_j}_0-\rtaui{i}\rVert\leq 20l_0, \\
            -\frac{\bm{v}^{t_j}_0}{\lVert\bm{v}^{t_j}_0\rVert}, & \lVert\bm{r}^{t_j}_0-\rtaui{i}\rVert\leq 10l_0,
        \end{cases}
    \end{equation*}
    where $\bm{r}^{t_j}_0\in\mathbf{Tr}$ is the closest point of the trail to $\rtaui{i}$.
    
    Finally, the speed limits and $\lak$ for the leader and the followers are given by:
    \begin{eqnarray*}
    &\lakle                        &= 2.0;\\
    &\text{V}_{\text{max,l}}       &= 4;\\
    &\lak                          &=
    \begin{cases}
    10, & \text{Flag}_f=0,\\
    \lakle, & \text{Flag}_f=1;\\
    \end{cases}\\
    &\text{V}_{\text{max,f}}       &=
    \begin{cases}
    l_0, & \text{Flag}_f=0,\\
    \text{V}_{\text{max,l}}, & \text{Flag}_f=1.\\
    \end{cases}
    \end{eqnarray*}
    
    \item {\bf The flocking and escaping of birds:} We set up the three dimensional region $[-80,80]\times[-80,80]\times[0,200]$. $N$ birds are randomly placed in the region of 
    
    \begin{equation*}
    \{(x,y,z)|(x+20)^2+(y-20)^2\leq30^2, 70<z<130\}.
    \end{equation*}
    
    We set their initial velocities to identically and independently follow the normal distribution with the mean of the $x$ direction being 7 and the mean of the $y$ direction being -7, as well as a standard deviation of 0.02. 
    
    Two predators are placed at the positions of $(-50,-50,100)$ and $(50,50,100)$. We do not assign randomness to the predators and just simply let them chase the bird that is closest to them at a fixed speed of 14. The positions of the predators at time $t$ is denoted by $\bm{r}_{p1}^t, \bm{r}_{p2}^t$.
    
    As mentioned in the main text, we use a combination of topological and metric methods to define the short-range interactions. We define $n_{i,top}^{t-\tau}$ as the collection of the nearest $n$ neighbors of the $i-$th bird at $t-\tau$, including the $i-$th bird itself.

    For the flock of the birds, we choose the following settings and simulate $M$ time steps.
    
    \begin{eqnarray*}
    &l_0     &= 1.0;\\
    &N       &= 1000;\\
    &M       &= 3000;\\
    &n       &= 10;\\
    &\tau    &= 0.03s;\\
    \end{eqnarray*}
    \begin{eqnarray*}
    &\lak    &= 10.0;\\
    &\lali   &= \lak;\\ 
    &\nalii  &= \{j\in n_{i,top}^{t-\tau}\left|\text{ } \lVert\rtaui{i}-\rtaui{j}\rVert\leq 20 l_0\}\right.;\\
    &g^{t,\tau}_{i,ali}  &= 1/\Nalii;\\
    &\larep  &= 5.0;\\
    &\nrepi  &= \{j\in n_{i,top}^{t-\tau}\left|\text{ } 0<\lVert\rtaui{i}-\rtaui{j}\rVert\leq 5 l_0\}\right.;\\
    &g_{rep} &= 1.0;\\
    &\laattr &= 5.0;\\
    &\nattri  &= \{j\left|\text{ } 30l_0<\lVert\rtaui{i}-\rtaui{j}\rVert\leq 80 l_0\}\right.;\\
    &g^{t,\tau}_{ratt}&=1/N;\\
    &\labo   &= 1.0;\\
    &\laex   &= 8.0.
    \end{eqnarray*}
    
    For the boundary effect, we consider two parts: in the $xy-$plane and in the $z-$direction. We denote $\rtauix{i},\rtauiy{i}, \rtauiz{i}$ and $\uboix,\uboiy, \uboiz$  as the $x,y,z$ components of $\rtaui{i}$ and $\uboi$, respectively. Then, we have
    
    \begin{eqnarray*}
    &D_b     &= 50;\\
    &L_b     &= 50;\\
    &U_b     &=150;\\
    &\nboixy   &= \left\{i\left|\text{ } \sqrt{(\rtauix{i})^2+(\rtauiy{i})^2}>D_b \right.\right\}; \\
    &g^{t,\tau}_{bou,xy} &= \sqrt{(\rtauix{i})^2+(\rtauiy{i})^2}-D_b;\\
    &\uboix &= -\rtauix{i}\left/\sqrt{(\rtauix{i})^2+(\rtauiy{i})^2};\right.\\
    &\uboiy &= -\rtauiy{i}\left/\sqrt{(\rtauix{i})^2+(\rtauiy{i})^2};\right.\\
    &\nboiz   &= \left\{i\left|\text{ } \rtauiz{i}<L_b \text{ or }  \rtauiz{i}>U_b\right.\right\}; \\
    &g^{t,\tau}_{bou,z}\uboiz &=
    \begin{cases}
    L_b-\rtauiz{i} &\text{when }\rtauiz{i}<L_b,\\
    U_b-\rtauiz{i} &\text{when }\rtauiz{i}>U_b.\\
    \end{cases}
    \end{eqnarray*}
    
    For the desire effect, we use the following setting to simulate the desire of escaping from the predators.
    
    \begin{eqnarray*}
    &D_d     &= 50;\\
    &\nexi   &= \left\{i\left|\text{ } \lVert\rtaui{i}-\bm{r}_{p}^{t-\tau}\rVert\leq D_d \right.\right\}; \\
    &g^{t,\tau}_{des}\uexti &=
    \begin{cases}
    \frac{10(\rtaui{i}-\bm{r}_{p}^{t-\tau})}{\lVert\rtaui{i}-\bm{r}_{p}^{t-\tau}\rVert^2}, &l_0<\lVert\rtaui{i}-\bm{r}_{p}^{t-\tau}\rVert\leq D_d;\\
    \frac{10}{l_0^2}(\rtaui{i}-\bm{r}_{p}^{t-\tau}), &\lVert\rtaui{i}-\bm{r}_{p}^{t-\tau}\rVert\leq l_0.\\
    \end{cases}
    \end{eqnarray*}
    
    Finally, we consider the lower and upper speed limit of the birds as $\text{V}_\text{min}=5$ and $\text{V}_\text{max}=15$.
    
\end{itemize}

\quad \\

With the above settings, we formulate the following algorithm for simulations, which has been implemented by writing the corresponding Python programs.

\begin{itemize}
    \item[S1.] Initialization based on the settings. In particular, define $t_i = i\tau, i=0,\dots, M$.
    \item[S2.] For $k=1,\dots, M$, do
    \begin{itemize}
        \item[S2.1.] Based on $\bm{s}_{t_{k-1}}$, compute the $\lambda,n,g,\bm{u}$ terms for each individual.
        \item[S2.2.] Then compute $\bm{F}^{t_{k-1}}_i$ based on equation (\ref{SIeqn:Fti}) for each individual.
        \item[S2.3.] Define $v_i=\lVert\bm{F}^{t_{k-1}}_i/\lak\rVert$. Then, the expected velocity of the $i-$th individual at time $t_k$ is given by 
        
        \begin{equation*}
        \bm{\mu}_i^{t_k}=\frac{\text{V}_\text{max}}{\max(\text{V}_\text{max},v_i)}\frac{\bm{F}^{t_{k-1}}_i}{\lak}. 
        \end{equation*}
        
        \item[S2.4.] If $\text{V}_\text{min}$ is also defined in the settings, then we multiply the above by $\text{V}_\text{min}/\min(\text{V}_\text{min},v_i)$, and use the result as $\bm{\mu}_i^{t_k}$.
        \item[S2.5.] Update $\bm{r}_i^{t_k}$ and $\bm{v}_i^{t_k}$ for each $i$ by
        \begin{eqnarray*}
            \bm{r}_i^{t_k} &=& \bm{r}_i^{t_{k-1}} + \bm{v}_i^{t_{k-1}}\tau,\\
            \bm{v}_i^{t_k} &=& \bm{\mu}_i^{t_k} +\bm{\epsilon_i}^{t_k},
        \end{eqnarray*}
        where $\bm{\epsilon_i}^{t_k}$ is the white noise term that follows a d-dimensional normal distribution with the mean of $\bm{0}$ and the covariance of $1/\lak\mathbf{I}$.
    \end{itemize}
    End do.
\end{itemize}

\section{\label{SIsec:NoDelay} The Model without Delay Effect} 

In Appendix \ref{SIsec:model}, we introduce the delay time $\tau$, which provides the following benefits. First, the dynamics are introduced into the framework. Second, the quadratic form of $\vt$ in (\ref{SIeqn:P1},\ref{SIeqn:Zt2}) is significantly simplified and Gaussian integral can be automatically introduced. That is because that setting the interaction effect to time $t-\tau$ as given information avoids the complexity of the corresponding quadratic matrix generated by the interaction neighbors (it doesn't matter whether the interaction is metric or topological within this setting).

In this section, we present the derivation and results of the framework at one given point of time $t$ without delay effect. Thus, the setting is static.

We don't, at least not directly, assign randomness to $\rt$ in our framework. In addition, $\urepi,\uattri,\uboi$, and $\uexti$ are only determined by $\mathbf{r}_{t-\tau}$. Since the mathematical challenges of setting $\tau=0$ are raised from the interactions of $\vt$. we assume that their corresponding $g$ terms are only determined by $\mathbf{r}_{t-\tau}$ as well to simplify the problem. It should be pointed out that the original dynamic framework does not require this assumption. In this section, we specify the letter $w\in\{rep, att, bou, des\}$, as the index of the effects that are determined by the positions only.

When $\tau$ is set to 0, the distribution (\ref{SIeqn:P1}) becomes
\begin{equation}
    \Ptt = \frac{1}{\Ztt}\exp\left[-\frac{1}{2}\vt\At\vt^T + \Ftt\vt^T\right], \label{SIeqn:Pttt}
\end{equation}
where $\Ftt$ is an $N-$dimensional ``vector'' whose $i-$th component is a $d-$dimensional vector given by
\begin{equation}
    \Fwti =  \sum_{w}\lwt{w}\Qwt{i}{w}, \label{SIeqn:FQ2}
\end{equation}
similar to (\ref{SIeqn:Fti}).

Without the loss of the generality, we assume $d=1$. For the cases of $d>1$, we only need to multiply all the  1-dimension probability density functions together as the final result, due to the format of $\Ptt,\vt$ and $\Ftt$.

 The $N\times N$ matrix $A_t$ is given by
\begin{equation}
    \At =  \lakt\bm{I} - \lalit\bm{N}_t, \label{SIeqn:A}
\end{equation}
where $\bm{N}_t=\{n_{ij}\}_{N\times N}$ is the alignment neighbor matrix. The element $n_{ij}$ is defined by
\begin{equation}
    n_{ij} = 
    \begin{cases}
        1, & \text{when } j\in\naliit \text{ and } j\neq i;\\
        0, & \text{otherwise}.
    \end{cases}
\end{equation}
An example of $\At$ can be found in FIG. \ref{SIfig:A}.

Recall that, in our original framework, $\nalii$ can be defined either topologically (the collection of the nearest $n$ neighbors of the $i-$th animal) or using metric (the collection of the individuals whose distances to the $i-$th animal are less than a given constant). The topological definition may introduce asymmetry to the model of (\ref{SIeqn:Pttt}). The fact that the $j-$th animal is one of the nearest $n$ neighbors of the $i-$th animal can not guarantee that the reverse is true. This asymmetry significantly increases the difficulty, sometimes even makes it impossible, to find the exact probability distribution. 

Thus, we only focus on the metric interaction, so that $\At$ is symmetric. In addition, we add the restriction of 
\begin{equation}
\lakt>\lalit\max(\text{eigenvalues of }\bm{N}_t),
\end{equation}
such that $\At$ is invertible. 

With the above restrictions, (\ref{SIeqn:Pttt}) can be solved as

\begin{equation}
    \Ptt = \frac{\exp\left[-\frac{1}{2}(\vt-\Ftt\At^{-1})\At(\vt-\Ftt\At^{-1})^T\right]}{\sqrt{(2\pi)^N/\det\At}}. \label{SIeqn:Pttt2}
\end{equation}
This is a multivariate normal distribution, the mean and covariance of which are given by
\begin{eqnarray}
\bm{\mu_{t}} &=& \Ftt\At^{-1}, \label{SImu22}\\
\bm{\Sigma_{t}} &=& \At^{-1} \label{SIvar22}.
\end{eqnarray}

We emphasize that $\tau$ is set to 0 in this section. $\Ftt$ and $\bm{\mu_t}$ are at the same time point. The equation (\ref{SImu22}) means that all the repulsion, attraction, boundary, and desire effects to all the individuals in the group will be assigned to every individual via $\At^{-1}$ simultaneously. However, this result is not consistent with reality. We demonstrate this conflict via the following example.

Consider a one-by-one queue of ants with a length of 5 meters. As shown by the individual 6, 7, and 8 in FIG. \ref{SIfig:A}, the matrix $\At$ should be a symmetric tridiagonal matrix as

\begin{equation*}
\At=
\begin{bmatrix}
\lambda_1    & -\lambda_2    & 0          & \cdots     & 0\\
-\lambda_2   & \lambda_1     & -\lambda_2 & \cdots     & 0\\
0            & -\lambda_2    & \lambda_1  & \cdots     & 0\\
\vdots       & \vdots        & \vdots     & \ddots     & \vdots    \\
0            & 0             & 0          & -\lambda_2 & \lambda_1\\
\end{bmatrix}_{N\times N}.
\end{equation*}
We use $\lambda_1=\lakt$ and $\lambda_2=\lalit$ here for simplification. The formula of $\At^{-1}$ can be found in \cite{ElMikkawy2006InvTridiag,Hu1996InvTridiag}. One can verify that the expected velocity of any individual $i$ will be affected by every individual in the queue. The ant at the end of the queue immediately knows that the ant at the head of the queue has encountered an obstacle, and react to it at the same time, even it is 5 meters away.

In summary, without the assumption of delay effect, this statistical framework may 
\begin{itemize}
    \item[1.] lose the ability to deal with dynamic cases;
    \item[2.] lose the ability to deal with topological interactions;
    \item[3.] cause a ``global  simultaneous  interaction''.
\end{itemize}

These problems may exist in other static maximum entropy models.

\setcounter{figure}{0}
\setcounter{table}{0}

\begin{figure}[tbhp]
\centering
\includegraphics[width=\linewidth]{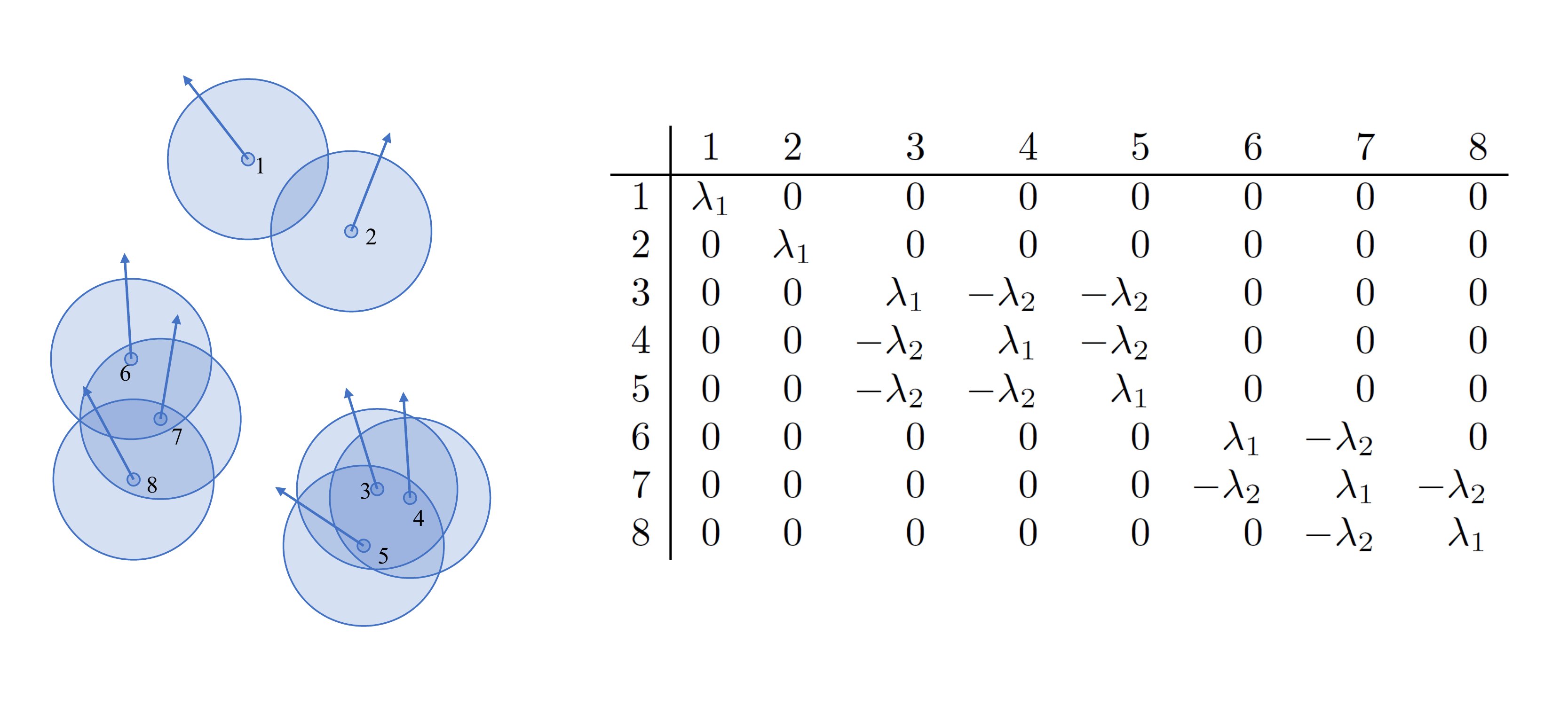}
\caption{An example of metric neighboring relations and the corresponding matrix $\At$. The light blue disc areas represent the neighborhood. We use $\lambda_1=\lakt$ and $\lambda_2=\lalit$ here for simplification.}
\label{SIfig:A}
\end{figure}

%